\documentclass[12pt]{article}

\usepackage[dvips]{graphicx}
\usepackage{latexsym}

\usepackage{amssymb,amsfonts,amsmath}
\usepackage{graphicx} 
\usepackage{indentfirst}

 \usepackage{bbm}

\parskip=\medskipamount

\def \un{\underline}
\newcommand {\cD}{{\cal D}}

\newcommand {\cG}{{\cal G}}

\newcommand {\cK}{{\cal K}}
\newcommand {\cL}{{\cal L}}
\newcommand {\cM}{{\cal M}}
\newcommand {\cN}{{\cal N}}

\newcommand {\cW}{{\cal W}}

\newcommand{\bW}{{\bf W}}

\def\a{\alpha}
\def \bi{\bibitem}
\def \ci{\cite}
\def\b{\beta}

\def\d{\delta}
\def\e{\epsilon}
\def\f{\phi}
\def\g{\gamma}
\def\G{\Gamma}

\def\m{\mu}
\def\n{\nu}

\def\p{\pi}
\def\q{\theta}
\def\r{\rho}
\def\s{\sigma}

\def\x{\xi}
\def\z{\zeta}
\def\D{\Delta}
\def\F{\Phi}
\def\J{\Psi}

\def\S{\Sigma}

\def\tr{{\rm tr}}
\def\rd{{\rm d}}
\def\ri{{\rm i}}

\newcommand{\ad}{{\dot{\alpha}}}                           

\newcommand{\pa}{\partial}                           
\newcommand{\hf}{\frac12}

\newcommand{\vf}{\varphi}
\newcommand{\sect}[1]{\setcounter{equation}{0}\section{#1}}

\newcommand{\be}{\begin{equation}}
\newcommand{\ee}{\end{equation}}
\newcommand{\bea}{\begin{eqnarray}}
\newcommand{\eea}{\end{eqnarray}}
\newcommand{\non}{\nonumber}
\newcommand{\bm}[1]{\mbox{\boldmath$#1$}}

\def\double #1{#1{\hbox{\kern-2pt $#1$}}}

\begin{document}

\begin{titlepage}

\begin{flushright}
hep-th/0703126\\
March, 2007\\
\end{flushright}
\vspace{5mm}

\begin{center}
{\Large \bf  Effective action of $\bm \b$-deformed $\bm {\cN = 4}$ SYM
theory: \\
Farewell to two-loop BPS diagrams
 }
\end{center}

\begin{center}

{\large  
Sergei M. Kuzenko\footnote{{kuzenko@cyllene.uwa.edu.au}}
and Ian N. McArthur\footnote{{mcarthur@physics.uwa.edu.au}}
} \\
\vspace{5mm}

\footnotesize{
{\it School of Physics M013, The University of Western Australia\\
35 Stirling Highway, Crawley W.A. 6009, Australia}}  
~\\

\vspace{2mm}

\end{center}
\vspace{5mm}

\begin{abstract}
\baselineskip=14pt
Within the background field approach, 
all two-loop sunset vacuum diagrams, which occur 
in the Coulomb branch of $\cN=2$ superconformal theories
(including $\cN=4$ SYM),
obey the BPS condition $m_3 = m_1 + m_2$, 
where the masses are generated by 
the scalars belonging to  a background $\cN=2$  vector multiplet. 
These diagrams can be evaluated exactly, 
and prove to be homogeneous quadratic functions of the 
one-loop tadpoles $J(m_1^2)$, $J(m_2^2)$ and $J(m_3^2)$, 
with the coefficients being rational functions 
of the squared masses. We demonstrate that,
if one switches  on the $\b$-deformation of the $\cN=4$ SYM 
theory, the BPS condition no longer holds,
and then generic two-loop sunset vacuum diagrams 
with three non-vanishing masses prove
 to be  characterized 
by the  following property: $ 2( m_1^2 m_2^2
+m_1^2 m_3^2 +m_2^2 m_3^2) > m_1^4 +m_2^4 +m_3^4$.
In the literature, there exist several techniques to
compute such diagrams.
${}$For the $\b$-deformed $\cN=4$ SYM theory, we carry out explicit
two-loop calculations of the K\"ahler potential and $F^4$ term.
Our considerations are restricted to the case of $\b$ real.
\end{abstract}
\vspace{1cm}

\vfill
\end{titlepage}

\newpage
\renewcommand{\thefootnote}{\arabic{footnote}}
\setcounter{footnote}{0}

\tableofcontents{}
\vspace{1cm}
\bigskip\hrule

\sect{Introduction}

In the family of finite $\cN=1$ supersymmetric theories
(see \cite{old,JJN} for an incomplete list of references),  
the exactly marginal  $\b$-deformation  \ci{LS} 
of the $\cN=4$ $SU(N)$  SYM theory has recently attracted 
some renewed attention, for it has been shown to possess
a supergravity dual description \cite{LM}.
In particular, 
in addition to stringy and non-perturbative aspects,
various field-theoretic properties
of the $\b$-deformed SYM have been studied
at the perturbative level, see
\cite{FG,PSZ,RSS,MPSZ,KT,MPPSZ,EMPS,RSS2,AKS} 
and references therein.
Naturally,  it is of special interest to understand what 
features of the $\cN=4$ SYM theory survive the deformation, 
as well as to determine new dynamical properties 
generated by the deformation. 
Of course, there are many non-trivial differences between 
the deformed and undeformed theories, and here we mention 
only a few of them.

Unlike the $\cN=4$ SYM theory, the finiteness condition 
in the deformed theory 
receives ``quantum corrections'' at different loop orders
\cite{FG,PSZ,RSS,JJN}.  This condition is  known exactly only 
for the real deformation in the large $N$ limit \cite{MPSZ}.
It is an exciting open problem to determine the exact condition
for  superconformal invariance at finite $N$.
We should point out that 
very interesting and conflicting results have appeared
regarding the fate of superconformal invariance 
for the complex deformation \cite{EMPS,RSS2}. Since a more detailed analysis 
of this issue is desirable, our consideration in this paper is restricted 
to the case of real $\b$.

As is well-known, in the Coulomb branch of general $\cN=2$ SYM theories, 
there are no quantum corrections to the effective K\"ahler potential 
beyond one loop, and no one-loop quantum corrections
in the $\cN=2$ superconformal models. 
In the $\b$-deformed $\cN=4$ SYM theory, 
however, one can expect the generation of a non-trivial 
superconformally invariant  K\"ahler potential 
at two and higher loops.
Similar holomorphic quantum corrections in the gauge sector are already
generated  at one loop  \cite{KT,DH}:
\be
 \frac{1}{16\p^2}  \,\int {\rm d}^2\q   \sum_{i<j}
\Big(W^i -W^j \Big)^2 \,
\ln \Bigg[ \frac{ g^2  (   \f^i -   \f^j    )^2}{ h^2\, ( q\, \f^i -  q^{-1} \, \f^j )
( q^{-1}\, \f^i -  q \, \f^j )    } 
\Bigg]~.
\ee
Here 
the $\cN=1$ chiral scalars $\F ={\rm diag} \,(\f^1, \dots ,  \f^N) $
and gauge-invariant  field strengths 
$ \cW_\a = {\rm diag} \,(W^1_\a , \dots ,W^N_\a) $
constitute a $\cN=2$ vector multiplet in the Cartan subalgebra 
of $SU(N)$, such that $\sum_{i=1}^{N} \f^i =  \sum_{i=1}^{N} W^i_\a =0$.
The above quantum correction disappears if the deformation parameters 
$q=\exp({\rm i}\p \b) $ and $h$ take the values $q=1$ and $h=g$ 
corresponding to the $\cN=4$ SYM theory, with $g$ the Yang-Mills
coupling constant.

The present paper, which is a continuation of \cite{KT}, 
is aimed at uncovering another interesting 
dynamical property of the $\b$-deformed theory that distinguishes
it from the $\cN=4$ SYM theory, and more generally from the 
$\cN=2$ superconformal models. 
It concerns the structure of two-loop sunset  diagrams with 
different masses, which have been 
studied by many groups including \cite{FJ,FJJ,DT,DST,CCLR}. 
We begin with a few general comments
about such Feynman diagrams. 

\begin{figure}[!htb]
\begin{center}
\includegraphics{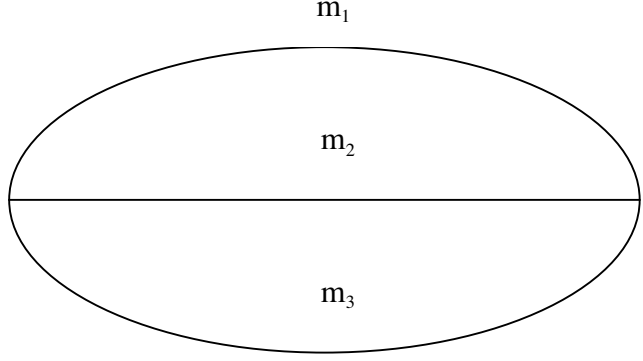}
\caption{Two-loop sunset diagram}
\end{center}
\end{figure}

When computing low energy effective actions 
within the background field formalism, 
one has to deal with two-loop sunset  vacuum integrals 
of the general form
\bea
I(\n_1,\n_2,\n_3; m_1^2,m_2^2,m_3^2) 
= \frac{(\m^2)^{4-d} }{(2\p)^{2d} }
\int \frac{ {\rm d}^d k \, {\rm d}^d q}
{ (k^2 +m_1^2)^{\n_1} (q^2 +m_2^2)^{\n_2}
((k+q)^2 +m_3^2)^{\n_3} }~,~~~
\label{n1n2n3}
\eea
with $\n_1$, $\n_2$ and $\n_3$ non-negative integers, 
see Fig. 1.
Davydychev and Tausk \cite{DT} 
derived recurrence relations that allow one to express 
$I(\n_1,\n_2,\n_3; m_1^2,m_2^2,m_3^2) $ in terms of
the master integral
$
I(1,1,1; m_1^2,m_2^2,m_3^2)  
$
and a product of one-loop tadpoles
\bea
J(m^2) = \frac{(\m^2)^{2-d/2} }{(2\p)^{d} }
\int \frac{ {\rm d}^d k}{k^2 +m^2}
= \frac{(\m^2/m^2)^{2-d/2} }{(4\p)^{d/2} } \, m^2
\,\G(1-d/2)~.
\label{tadpole}
\eea
The recurrence relations obtained in \cite{DT} are:
\bea 
&& {\mathbb I}(\n_1 +1,\n_2, \n_3) 
= -\,\frac{1}{\n_1 m_1^2 \D (m_1^2,m_2^2,m_3^2) }
\Big\{ {\mathbb I}(\n_1, \n_2, \n_3)\Big[
\n_2 (m_1^2 -m_3^2)(m_1^2 -m_2^2 +m_3^2) 
\non \\
&& \quad +   \n_3 (m_1^2 -m_2^2)(m_1^2 +m_2^2 -m_3^2) 
+ d\, m_1^2  (m_1^2 -m_2^2 +m_3^2) -\n_1   \D (m_1^2,m_2^2,m_3^2) \Big]
\non \\
&& \quad + \n_2 m_2^2 (m_1^2 -m_2^2 +m_3^2) 
\Big[ {\mathbb I}(\n_1, \n_2 +1, \n_3 -1) -{\mathbb I}(\n_1-1, \n_2 +1, \n_3) \Big] 
\non \\
&& \quad + \n_3 m_3^2 (m_1^2 +m_2^2 -m_3^2) 
\Big[ {\mathbb I}(\n_1, \n_2 -1, \n_3 +1) -{\mathbb I}(\n_1-1, \n_2 , \n_3+1) \Big] \Big\}~, 
\label{recurrence}
\eea
where we have used the condensed notation
$ {\mathbb I}(\n_1,\n_2,\n_3) \equiv
I(\n_1,\n_2,\n_3; m_1^2,m_2^2,m_3^2) $, and 
\be
\D (m_1^2,m_2^2,m_3^2) = 2( m_1^2 m_2^2
+m_1^2 m_3^2 +m_2^2 m_3^2) -( m_1^4 +m_2^4 +m_3^4)~.
\label{Delta}
\ee
The master integral $I(1,1,1; m_1^2,m_2^2,m_3^2) $
can be evaluated 
using the techniques developed, e.g.,  in  \cite{FJJ,DT,CCLR}.

On the Coulomb branch of $\cN=2$ superconformal theories, 
we have $U(1)$ charge conservation at each vertex of the supergraphs. 
In particular, for the two-loop sunset diagrams we get
\be
e_1 + e_2 + e_3 =0~.
\ee
Because of the BPS condition $m_i = Z |e_i|$, 
the requirement of charge conservation implies
\be
  m_1 = m_2 + m_3~, \quad {\rm or} \quad ~
m_2 = m_1 +m_3 ~, \quad {\rm or} \quad ~
m_3 = m_1 +m_2~.
\ee 
This leads to the condition
\be
\D (m_1^2,m_2^2,m_3^2) = 0~
\label{D0}
\ee
in arbitrary  $\cN=2$ superconformal theories, due 
to the factorization property \cite{T,CCLR}
\be
\frac {\D (m_1^2,m_2^2,m_3^2) }{ m_1 +m_2 +m_3 }=
(m_1 +m_2 -m_3) (m_1-m_2 +m_3)
(-m_1 +m_2 + m_3  )~.
\ee 
As a result, the  recurrence relations
(\ref{recurrence}) cannot be applied.
Also, because of (\ref{D0}), 
we cannot generate integrals 
(\ref{n1n2n3}) with arbitrary $\n_i$
from $I(1,1,1; m_1^2,m_2^2,m_3^2) $
by differentiation with respect to $m_i^2$.
In other words, special consideration is
required in order to compute momentum integrals 
(\ref{n1n2n3}) under the BPS condition
(\ref{D0}).

Actually, recurrence relations for the case (\ref{D0}) have been found 
by Tarasov \cite{T}. They imply that 
all  BPS integrals (\ref{n1n2n3}) are given in terms of elementary functions, 
unlike the generic case $\D (m_1^2,m_2^2,m_3^2) \neq 0 $,
where integral representations for $I(1,1,1; m_1^2,m_2^2,m_3^2) $
involve transcendental functions
 \cite{FJJ,DT,CCLR}.
More precisely, the BPS integrals are 
homogeneous quadratic functions of the 
one-loop tadpoles $J(m_1^2)$, $J(m_2^2)$ and $J(m_3^2)$, 
with the coefficients being rational functions 
of the squared masses. 
In the present paper, we will re-derive this result using several different 
approaches.

If one switches from the $\cN=4$ SYM theory to its $\b$-deformation, 
it turns out that  the BPS condition (\ref{D0}) no longer holds. 
As is shown below, with $\b$ real, one typically has
$\D (m_1^2,m_2^2,m_3^2) >0$ if all masses are non-vanishing.
Two-loop calculations of the effective action become much more involved, 
as compared with the $\cN=4$ case, and one has to use the full power of 
the techniques developed in  \cite{FJJ,DT,CCLR}.

This paper is organized as follows. 
In section 2 we give a detailed study of  integrals (\ref{n1n2n3}) 
under the BPS condition (\ref{D0}).
Section 3 is devoted to various aspects of the 
background field quantization of the $\b$-deformed $\cN=4$ 
SYM theory, including the specification of the background superfields
chosen. In section 4 the two-loop contributions to the effective action 
are decomposed into a set of terms involving only $U(1)$ Green's functions.
Exact covariant superpropagators are given in section 5.
The two-loop quantum corrections are evaluated in section 6.
Two technical appendices are included at the end of the paper. 
Appendix A reviews and elaborates on the approach developed in \cite{FJJ}.
Appendix B contains the $SU(N)$ conventions adopted in this paper.

\sect{Two-loop BPS integrals}

Consider a vacuum two-loop  integral 
$ I(\n_1 ,\n_2 , \n_3 ; x,y,z)$ as in (\ref{n1n2n3}), 
with $\n_1, \n_2$ and $\n_3$ non-negative integers, 
and $(x,y,z)=(m_1^2, m_2^2,m_3^2)$.
Under the BPS condition $\D(x,y,z)=0$, 
this integral turns out to be a homogeneous 
quadratic function of $J(x)$, $J(y)$ and $J(z)$, 
with the coefficients being rational functions 
of $x$, $y$ and $z$. 
The simplest way to establish 
this result is   
by making use of  a differential equation for 
the completely symmetric master integral $I(x,y,z) = I(1 ,1, 1 ; x,y,z)$, 
defined in eq. (\ref{I(xyz)}), that was 
presented in \cite{CCLR}.

Using eq. (\ref{pde1}) and 
two more equations that follow 
from (\ref{pde1}) by applying cyclic permutations 
of $x$, $y$ and $z$,  
one can  deduce   
a differential equation involving a single partial derivative of $I$, 
say with respect to $z$ \cite{CCLR}. It reads
\bea
&&\D(x,y,z)\, \frac{\pa}{\pa z} I(x,y,z)= (d-3) (x+y-z) I(x,y,z) \non \\
&&\qquad
 +\frac{2-d}{2} \Big\{ \frac{-x+y+z}{z} J(x)J(z) 
+\frac{x-y+z}{z} J(y)J(z) -2J(x)J(y) \Big\}~,
\label{BPSmaster}
\eea
with $\D(x,y,z)$ defined in (\ref{Delta}).
Implementing
cyclic permutations 
of $x$, $y$ and $z$,  
one generates two more equivalent equations.

Now, let us apply an operator
$$
\frac{\pa^{\n_1+\n_2+\n_3}}{\pa x^{\n_1} \pa y^{\n_2} \pa z^{\n_3}}
$$ 
to both sides of eq. (\ref{BPSmaster}), and 
in the end impose the BPS condition
 $\D(x,y,z)=0$. The latter implies that the term of highest order in derivatives 
of $I$ drops out.

Using the condensed notation 
${\mathbb I}(\n_1 ,\n_2 , \n_3 )\equiv I(\n_1 ,\n_2 , \n_3 ; x,y,z)$, 
on the two-dimensional surface  $\D(x,y,z)=0$
we obtain the following algebraic equations: 
\bea
&&\n_1 \frac{\pa \D}{\pa x} {\mathbb I}(\n_1 ,1+\n_2 , 2+\n_3 )\, (\n_1-1)! \n_2!(\n_3+1)! 
\non \\
&&\quad
+\n_2 \frac{\pa \D}{\pa y} {\mathbb I}(1+\n_1 ,\n_2 , 2+\n_3 )\, \n_1! (\n_2-1)!(\n_3+1)! \non \\
&&\quad +\hf (2\n_3-d+3)  \frac{\pa \D}{\pa z} 
{\mathbb I}(1+\n_1 ,1+\n_2 , 1+\n_3 )\, \n_1! \n_2!\n_3! \non \\
&&\quad +\n_3(\n_3-d+2) {\mathbb I}(1+\n_1 ,1+\n_2 , \n_3 )\, \n_1! \n_2!(\n_3-1)! \non\\
&&\quad +
\n_1(\n_1-1) {\mathbb I}(\n_1 -1,1+\n_2 , 2+\n_3 )\, (\n_1-2)! \n_2!(\n_3+1)! \non \\
&&\quad+\n_2(\n_2-1) {\mathbb I}(1+\n_1 ,\n_2 -1, 2+\n_3 )\, \n_1! (\n_2-2)! (\n_3+1)! \non \\
&&\quad- 2\n_1 \n_2   {\mathbb I}(\n_1 ,\n_2 , 2+\n_3 )\, (\n_1-1)! (\n_2-1)! (\n_3+1)! \non\\
&&\quad+(d-3-2 \n_3) \Big\{
\n_1 {\mathbb I}(\n_1 ,1+\n_2 , 1+\n_3 )\, (\n_1-1)! \n_2!  \non \\
&&\qquad \qquad \qquad \qquad 
+\n_2 {\mathbb I}(1+\n_1 ,\n_2 , 1+\n_3 )\, \n_1! (\n_2-1)!  \Big\} \n_3!
\non\\
&&=  (-1)^{\n_1+\n_2+\n_3} \frac{2-d}{2}\frac{\pa^{\n_1+\n_2+\n_3}}
{\pa x^{\n_1} \pa y^{\n_2} \pa z^{\n_3}}
 \Big\{ \frac{-x+y+z}{z} J(x)J(z) 
+\frac{x-y+z}{z} J(y)J(z) \non \\
&& \qquad \qquad \qquad \qquad \qquad \qquad \qquad 
-2J(x)J(y) \Big\}\Big|_{\D(x,y,z)=0}~.
\label{BPSmaster4}
\eea
Eq. (\ref{BPSmaster4}) constitutes  recurrence relations 
allowing one to compute, for even $d$, arbitrary two-loop  integrals $I(\n_1 ,\n_2 , \n_3 ; x,y,z)$, 
with $\n_1,\n_2,\n_3 =1,2, \dots $, under the BPS condition  $\D(x,y,z)=0$
and the assumption that all three masses are non-vanishing.\footnote{If
one of the masses vanishes, the BPS condition implies 
that the other two masses are equal, and then the integrals are evaluated 
by elementary methods.}   
Choosing $\n_1=\n_2=\n_3=0$ in (\ref{BPSmaster4}) gives \cite{FJJ,CCLR,EZ}
\bea
&&I_{\rm BPS}(x,y,z) 
=
\frac{d-2}{2(d-3)} \,\frac{1}{\pa_z\D} \Big\{ 
\frac{\pa_x \D}{z} J(x)J(z) 
+\frac{\pa_y \D}{z} J(y)J(z) 
-4J(x)J(y) \Big\}
\non \\
\non \\
&=& -\frac{d-2}{4(d-3)} 
\Big\{ \frac{x-y+z}{xz} J(x)J(z) 
+ \frac{-x+y+z}{yz} J(y)J(z) 
+ \frac{x+y-z}{xy} J(x)J(y) \Big\}~. ~~~~~~~~
\label{BPS000}
\eea
Here we have used 
the BPS restrictions of the identities
\bea
\frac{1}{4} \pa_x\D \,\pa_y\D = \D -z\pa_z \D~,
\quad
\frac{1}{4} \pa_x\D \,\pa_z\D = \D -y\pa_y \D~,
\quad 
\frac{1}{4} \pa_y\D \,\pa_z\D = \D -x\pa_x \D~~~~
\label{Iden1}
\eea
and
\bea
\frac{1}{4}\big( \pa_x \D \big)^2  +  \D = 4yz~, \quad 
\frac{1}{4}\big( \pa_y \D \big)^2  +  \D = 4xz~, \quad 
\frac{1}{4}\big( \pa_z \D \big)^2  +  \D = 4xy~,
\label{Iden2}
\eea
to bring $I(x,y,z)$ to a manifestly symmetric form 
with respect to $x,$ $y$ and $z$.  
As is seen from these identities, the combitations
$\pa_x\D$, $\pa_y\D$ and $\pa_z\D$ are non-zero at $\D =0$. 
In addition, two of them are  positive and the third  negative.
Indeed, without loss of generality, we can choose $x=m_1^2$, 
$y=m_2^2$ and $z=(m_1+m_2)^2$, so that
for the combinations $ \pa_x \D = 2(-x+y+z)$, 
$\pa_y \D = 2(x-y+z)$ and  $\pa_z \D = 2(x+y-z)$ we get
\bea
 \pa_x \D 
= 4m_2 (m_1 +m_2)~, \quad 
\pa_y \D 
= 4m_1(m_1+m_2)~, \quad
 \pa_z \D 
= -4m_1m_2 ~.
\label{BPSder}
\eea

As a less trivial example,
choosing $\n_1=\n_2=0$ and $ \n_3=1$ in (\ref{BPSmaster4}) gives
\bea
 I_{\rm BPS}(1,1,2;x,y,z)&=& \frac{d-2}{4(5-d) } \frac{1}{xyz^2} 
\Big[ y \Big\{ z-\hf (d-4) (x-y+z)\Big\} J(x)J(z) \non \\
&&  + x \Big\{ z-\hf (d-4) (-x+y+z)\Big\} J(y)J(z)
-z^2 J(x) J(y)\Big]~.~~~
\label{I(1,1,2)}
\eea
More generally, choosing $\n_1=\n_2=0$ and $ \n_3>0$ in (\ref{BPSmaster4}) gives
the recurrence relation:
\bea
&& I_{\rm BPS}(1,1,1+\n ;x,y,z)= \frac{1}{(2\n+3-d)(x+y-z) }  
\Bigg[ (d-2-\n) I(1,1,\n ;x,y,z) \non \\
& & \qquad \qquad + \frac{(-1)^\n}{\n! \,2^{\n+1}} (d-2)(d-4)\dots (d-2\n) 
\,\frac{J(z)}{z^{\n+1}}
\non  \\
&& \qquad \qquad  \times  \Big\{ (d-2-2\n)(x-y)\big(J(x)- J(y)\big)
-(d-2)z\big(J(x) +J(y)\big) \Big\} \Bigg]~. ~~~~
\eea

The recurrence relations (\ref{BPSmaster4}) look somewhat messy, 
 although their derivation is completely trivial.
More elegant recurrence relations, albeit equivalent to eq. (\ref{BPSmaster4}), 
were derived in \cite{T}.
In both cases, the recurrence relations clearly demonstrate that the BPS integrals
are homogeneous quadratic functions of the 
one-loop tadpoles $J(x)$, $J(y)$ and $J(z)$, 
with the coefficients being rational functions 
of the squared masses. 

In practical terms, it may be simpler to compute 
the  two-loop  BPS integrals 
by first differentiating the expression in eq. (\ref{final}) 
for the master integral
with respect to $x$, $y$ and $z$, 
and then implementing the limit $\D \to 0$. This requires certain care, 
since the limit $\D \to 0$ is actually singular (one should also make use
of the fact that, among the combinations $\pa_x \D$, $\pa_y \D$
and $\pa_z \D$, two are positive, and the third negative, as eq. 
(\ref{BPSder}) explicitly shows). 
As an example, let us evaluate $I(1 ,1 , 2 ; x,y,z)|_{\D=0} $ using 
the functional representation (\ref{final}). Differentiating  the right-hand side
of (\ref{final}) with respect to $z$ and then setting $\D=0$ gives
\bea
I_{\rm BPS}(1 ,1 , 2 ; x,y,z)= 
\frac{\G'}{2(1+2\e)}  \Bigg\{ (xy)^{-\e} 
&-& (xz)^{-\e} \Big[ 1+ 2\e +2 \e \, \frac{\pa_z \D}{\pa_y \D} \Big] \non \\ 
&-& (yz)^{-\e} \Big[ 1+ 2\e +2 \e \, \frac{\pa_z \D}{\pa_x \D} \Big] \Bigg\}~,
\eea
with $\G'$ given in eq. (\ref{Gamma'}).
This can be seen to agree with the representation (\ref{I(1,1,2)}) if one makes use 
of (\ref{BPSder}).

The powerful techniques developed in  \cite{FJJ}  and \cite{DT} to compute 
the two-loop master integral  $I(x,y,z)$, eq. (\ref{I(xyz)}), 
are quite involved.
Remarkably, the first-order inhomogeneous ODE (\ref{BPSmaster})  
and the value of $I(x,y,z)$ at $\D=0$, eq. (\ref{BPS000}), 
comprise all the ingredients one needs to compute $I(x,y,z)$ by elementary means 
\cite{CCLR}, in the case with three non-vanishing masses.
Let us consider, for definiteness, the domain $\D(x,y,z)>0$.
Eq. (\ref{BPSmaster})  is integrated as follows:
\bea
I(x,y,z) &=&-\D^{(d-3)/2}(x,y,z) \, \Big\{ \int\limits_{z_0}^{z} {\rm d}t \,
\frac{\J(x,y,t)}{\D^{(d-1)/2}(x,y,t)}
+\frac{ I(x,y,z_0) }{\D^{(d-3)/2}(x,y,z_0)} \Big\}~, 
\label{repa1}\\
\J (x,y,z) &=& \hf \Big\{ \pa_x\D \, J(x) J'(z) + \pa_y \D \, J(y) J'(z) 
-2(d-2) J(x)J(y)\Big\}~, \non
\eea
for some $z_0$ such that $\D(x,y,z_0)>0$.
Since the right-hand side of (\ref{repa1}) does not depend on 
$z_0$, we can consider the limit $z_0 \to z_{\rm BPS} = (m_1 +m_2)^2$, 
with the latter point such that
$\D(x,y, z_{\rm BPS})=0$ and $I(x,y,z_{\rm BPS}) 
= I_{\rm BPS}(x,y,z_{\rm BPS})$.
In this limit, however, the integral in  (\ref{repa1}) becomes singular at the lower limit. 
To get rid of singularities, we can first integrate by parts in  (\ref{repa1}) by making use 
of the third identity in  (\ref{Iden2}). More precisely, using the identity 
\bea
\frac{1}{\D^{n+1}} = \frac{1}{4xy} \,\Big\{ \frac{1}{\D^{n}} 
-\frac{\pa_t \D}{4n} \,\pa_t \Big( \frac{1}{\D^{n}} \Big) \Big\} ~,
\qquad \D=\D(x,y,t)
\label{Dpower}
\eea
on the right of  (\ref{repa1}) in order to integrate by parts, and then 
implementing the limit $z_0 \to z_{\rm BPS}$, we obtain 
\bea
I(x,y,z) &=&-
\frac{  \D^{(d-3)/2}(x,y,z) }{4(d-3)xy}\, 
\int\limits_{z_{\rm BPS} }^{z} {\rm d}t \,
\frac{\J_1(x,y,t)}{\D^{(d-3)/2}(x,y,t)}
+\frac{ \J(x,y,z) \,\pa_z\D(x,y,z) }{8(d-3)xy} 
~, ~~~~~
\label{repa2}\\
\J_1 (x,y,z) &=&
(d-4)\J +\hf \pa_z\D \,\pa_z \J~. \non
 \eea
The first term on the right of (\ref{repa2}) can  again 
be integrated by parts, using the identity (\ref{Dpower}),
and so on. 
This can be seen to generate 
a representation for 
$I(x,y,z) $ 
as a  series 
in powers of $\D$.
In particular, for any positive integer $k$ we have
\bea
I(x,y,z) &=&- \frac{  \D^{(d-3)/2}(x,y,z) }{(d-3)(d-5)\cdots (d-3-2k)}
\Big(\frac{1}{4xy}\Big)^{k+1}\, 
\int\limits_{z_{\rm BPS} }^{z} {\rm d}t \,
\frac{\J_{k+1}(x,y,t)}{\D^{(d-3-2k)/2}(x,y,t)} \non \\
&&+ \frac{\pa_z \D(x,y,z)}{8xy}\sum_{p=0}^{k}
\Big(\frac{\D(x,y,z)}{4xy}\Big)^{p} 
\frac{\J_p (x,y,z)}{(d-3)(d-5)\cdots (d-3-2p)}~,
\label{repa4}
\eea
with 
\bea
\J_{k+1}= (d-4-2k) \J_k +\hf \pa_z \D \, \pa_z \J_k~,
\qquad \J_0 =\J~.
\eea
This representation is useful, e.g., for an alternative evaluation 
of the BPS integrals.

To conclude this section, we note that 
an alternative solution to equation (\ref{BPSmaster}) 
was given in \cite{CCLR} in the context of the epsilon-expansion.

\sect{The $\bm \beta$-deformed $\bm \cN$ = 4 SYM theory}

The $\b$-deformed $\cN=4$ 
$SU(N)$ SYM theory 
is described by the action 
\bea
S &=& 
\int {\rm d}^8 z \, {\rm tr}\,( \F_i^\dagger \,\F_i )
+   \frac{1}{g^2} \int {\rm d}^6 z \, 
{\rm tr} (
\cW^\a \cW_\a ) \non \\
&&+ \Big\{
h  \int {\rm d}^6 z \, {\rm tr} (
q\, \F_1 \F_2 \F_3 
-q^{-1}\,
\F_1 \F_3 \F_2 )
~+~ {\rm c.c.} \Big\}~, \ \ \ \ \  q \equiv 
{\rm e}^{{\rm i} \p \b }\ ,
\label{b-deformed-N=4SYM}
\eea
where 
$q $ is the deformation parameter, $g$ is the gauge coupling constant, 
and $h$ is related to $g$ and $q$  by the 
condition of quantum conformal invariance.
The latter is not yet known exactly, since it is expected to receive 
quantum corrections at arbitrary  loop orders, and the higher loop 
corrections are hard to evaluate in closed form.\footnote{In 
the large  $N$ limit,  the condition of finiteness  
 (\ref{finite2}) becomes $|h| =g$, as in the $\cN=4$ theory.
In this limit, it was argued in  \cite{MPSZ}
using the analogy \cite{LM} with  the  non-commutative theory,  
that this is actually  the exact condition for conformal invariance to all loops.
The case of complex $\b$ is more subtle.}  
To  {\it two-loop} order, the condition of quantum conformal invariance
for real $\b$ is as follows \cite{FG,PSZ,JJN} (see also \cite{NP}): 
\be
|h|^2 \, \Big( 1 -{1\over N^2}\, \Big|q  - {1\over q} \Big|^2 \Big) =g^2~, 
\qquad \quad |q|=1~.
\label{finite2}
\ee
The original $\cN=4$  theory corresponds to $|h|=g$ and $q=1$.  
In what follows, we restrict our consideration to the case 
of real $\b$.

It is useful to view the $\cN=1$ supersymmetric theory with 
action (\ref{b-deformed-N=4SYM}) as a pure $\cN=2$ 
super Yang-Mills theory (described by $\F_1$ and $\cW_\a$)
coupled to a {\it deformed  hypermutiplet} in the adjoint 
(described by $\F_2$ and $\F_3$). Here we are interested specifically 
in the quantum  effects induced by the deformation. 
Since the deformation occurs only 
in the hypermultiplet sector, our analysis of the effective action 
will concentrate on evaluating the two-loop quantum corrections
from all  the supergraphs involving quantum hypermultiplets.

The extrema of the scalar potential 
generated by   (\ref{b-deformed-N=4SYM})
are  described by the equations 
(here ${\bm \F}_i$ denote the first components of the chiral superfields $\F_i$)
\be
 \sum_i [ {\bm \F}_i  \, ,{\bm \F}_i{}^\dagger  ] =0 ~, 
\qquad 
q \, {\bm \F}_i {\bm \F}_{i+1} -q^{-1} \,{\bm \F}_{i+1} {\bm \F}_i 
= {1\over N}  \, (q - q^{-1} ) \,{\rm Tr} ({\bm \F}_i {\bm \F}_{i+1} )
\, \mathbbm{1} ~.
\ee
In what follows, 
we shall consider the simplest 
special  solution 
 \be 
{\bm \F}_1 \equiv {\bm \F} \ , \ \ \ \ \ \ \ 
   {\bm \F}_2={\bm \F}_3=0 \ , \ee 
where $\bm \F$ is a  diagonal traceless $N \times N$ matrix.
This solution is especially interesting in the context of 
quantum $\cN=2$ super Yang-Mills theories, for it corresponds
to the Coulomb branch.

To quantize the theory, we 
use the $\cN=1$ 
background field formulation \cite{GGRS} and 
split the dynamical variables into  background 
and quantum,
\bea
  \F_i ~ \to ~ \F_i +\vf_i ~, 
\qquad   \quad
\cD_\a ~ \to ~ {\rm e}^{-g\,v} \, \cD_\a \, {\rm e}^{g\, v}~, 
\qquad
{\bar \cD}_\ad ~ \to ~ {\bar \cD}_\ad~,
\label{bq-splitting}
\eea
with lower-case letters used for 
the quantum superfields. 
Then the action becomes 
\bea
S= 
\int {\rm d}^8 z \, {\rm tr} \Big(
( \F_i  +\vf_i{} )^\dagger \,
{\rm e}^{g\,v} \, ( \F_i +\vf_i) \, {\rm e}^{-g\, v} \Big)
&+&  {1\over g^2}\int {\rm d}^6 z \, {\rm tr} \Big( \bW^\a \bW_\a  \Big) \non \\
+ \Big\{
\int {\rm d}^6 z \, \cL_{\rm c} (\F_i + \vf_i) 
&+& {\rm c.c.} \Big\}~,
\label{back-quant-split}
\eea
where
$\cL_{\rm c}(\F_i)$ stands for the superpotential 
in   (\ref{b-deformed-N=4SYM}), and
\bea
\bW_\a &=& - {1\over 8} {\bar \cD}^2 \Big( 
{\rm e}^{-g\,v}\, \cD_\a \,{\rm e}^{g\,v} \cdot 1 \Big)  = \cW_\a
- {1\over 8} {\bar \cD}^2 \Big( 
g\, \cD_\a v- \hf \, g^2[v, \cD_\a v] 
\Big) + O(v^3)~.  
\eea 
We choose  $\F_2=\F_3=0$ and $\F_1 \equiv \F \neq 0$.
Since both the gauge and matter background superfields are 
non-zero, it is convenient to use the 
$\cN=1$ supersymmetric 't Hooft gauge 
(a special case of the supersymmetric $R_\x$-gauge
introduced in \cite{OW} and further developed in \cite{BBP}),
following the technical steps described in detail in Refs. \cite{KM3,KM4,KT}.

Modulo ghost contributions,
the  quadratic part, $S^{(2)}$,  of the gauge-fixed action  
can be shown to include two terms corresponding, respectively, 
to the pure $\cN=2$ SYM sector ($ S^{(2)}_{\rm I} $) and 
to the deformed hypermultipet ($S^{(2)}_{\rm II}$). They are: 
\bea 
S^{(2)}_{\rm I} 
 &=&  - \frac{1}{2}
\int {\rm d}^8 z \,{\rm tr}\,
\Big( v \,\Box_{\rm v}  v - g^2 \,v\,[\F^\dagger, [\F, v]] \Big) 
\non \\
&&+ \int {\rm d}^8 z \,
{\rm tr}\,\Big( \vf_1^\dagger \,\vf_1 
- g^2 \,[\F^\dagger , [\F, \vf_1^\dagger ]] \, 
(\Box_+)^{-1}\, \vf_1 \Big) 
~+~ \dots  ~;
\label{quad-prel}
\\
S^{(2)}_{\rm II} &=&
 \int {\rm d}^8 z \,{\rm tr}\,
 \Big( 
\vf_2^\dagger \, \vf_2 
+  \vf_3^\dagger \,  \vf_3  
 \Big) 
+ \int {\rm d}^6 z \, {\rm tr}\, 
 \vf_3 \cM_{(h,q)} \vf_2 
+ \int {\rm d}^6 {\bar z} \, {\rm tr}\,
 \vf_2^\dagger \cM_{(h,q)}^\dagger  \vf_3^\dagger 
~,~~~
\label{quad-prel2}
\eea 
where the mass operator $\cM_{(h,q)}$ 
and it Hermitian conjugate  $\cM_{(h,q)}^\dagger$ are defined by
their action on a Lie-algebra valued superfield: 
\bea
\cM_{(h,q)} \S &=& h \,(q \,\F\, \S 
- {1\over q} \,\S \,\F) -  {h\over N}  \,(q  - {1\over q} )\,
{\rm tr} ( \F\, \S ) \, \mathbbm{1} ~, \non \\
\cM_{(h,q)}^\dagger \S &=& {\bar h} \,(\frac{1}{q} \,\F^\dagger \S 
- q \,\S \,\F^\dagger) +  {{\bar h}\over N}  \,(q  - {1\over q} )\,
{\rm tr} ( \F^\dagger \S ) \, \mathbbm{1} ~,
\label{massop} 
\eea
such that 
\bea
 \cM_{(h,q)}^{\rm T} = - \cM_{(h,{1 \over q})} =\cM_{(h,-{1\over q})} ~.
\eea  
In the expression for $S^{(2)}_{\rm I}$,
 the dots stand for the terms with derivatives
of the background (anti)chiral superfields 
 $\F^\dagger$ and $\F$.
The second-order operators $\Box_{\rm v}$ and
$\Box_+$ in  (\ref{quad-prel})
denote the vector and
the covariantly chiral  d'Alembertians, respectively. 
\bea 
{\Box}_{\rm v} 
&=& \cD^a \cD_a - \cW^\a \cD_\a +{\bar \cW}_\ad {\bar \cD}^\ad ~,
\qquad 
\Box_+ = \cD^a \cD_a - \cW^\a \cD_\a -\hf \, (\cD^\a \cW_\a)~.
\eea

${}$From (\ref{back-quant-split}) we can read off 
the cubic and quartic hypermultiplet vertices  
which generate the two-loop supergraphs of interest. 
The cubic vertices are:
\bea 
S^{(3)}_{\rm I} &=& g
 \int {\rm d}^8 z \,  {\rm tr}
\Big( \vf_2^\dagger \, [v ,\vf_2 ]
+  \vf_3^\dagger \, 
[v,  \vf_3 ] \Big)~;
\label{hyper-vector}
\\
S^{(3)}_{\rm II}
 &=& h \int \rd^6 z \, \tr  \Big( q \, \vf_1  \vf_2  
\vf_3 - q^{-1} \, \vf_1  \vf_3  \vf_2 \Big) + {\rm c.c. } 
\label{hyper-3}
\end{eqnarray}
${}$Finally, we should take into account the quartic 
hypermultiplet vertices
\bea
S^{(4)} &=& \hf g^2 \int {\rm d}^8 z \, {\rm tr}
\Big( \vf_2^\dagger \, [v,[v ,\vf_2 ] ]
+  \vf_3^\dagger \, 
[v,[v,  \vf_3 ] ]\Big)~.
\label{hyper-4}
\eea

It is convenient to introduce the following ``deformation'' 
of the generators in the adjoint representation:
\be
\left({T}_{(h,q)}^a\right)^{bc} = - \, \frac{\ri }{2} \,h \,( q + q^{-1}) \,f^{abc} 
- \hf h (q - q^{-1}) d^{abc},
\label{defT}
\ee
with the algebraic properties
\be
 \left({T}_{(h,q)}^a\right)^{\rm T}
=- {T}_{(h,{1\over q})}^a ~, \qquad
\left( {T}_{(h,q)}^a\right)^\dagger
=  {T}_{( {\bar h},{1\over q})}^a~.
\ee
In the limit that the deformation vanishes, these reduce to the generators in 
the adjoint representation, multiplied by the coupling constant $g.$ 
Using this notation, the cubic vertex (\ref{hyper-3}) takes the form
\be
S^{(3)}_{\rm II} = -  \int \rd^6 z 
\left( {T}_{(h,q)}^a \right)^{bc} 
 \vf_1^a \vf_2{}^b \, \vf_3{}^c 
+  \int \rd^6 \bar{z} \,
\big(  {T}_{({\bar h}, {1\over q})}^a \big)^{bc}
\bar{\vf}_1^a \, \bar{\vf_2}^b \, \bar{\vf_3}^c ~.
\ee

In what follows, 
the background superfields will be  chosen  
to satisfy the following on-shell conditions:
\be
[\F , \F^\dagger ] = 0~, \qquad \cD_\a \F =0~,
\qquad  \cD^\a \cW_\a = 0~, 
\label{back-con-1}
\ee
with some  additional conditions on the background
superfields to be imposed later on. Then, 
the Feynman propagators 
for the actions (\ref{quad-prel}) and (\ref{quad-prel2}) 
can be expressed in terms of two Green's functions 
in the adjoint representation, $\stackrel{\hookrightarrow}{G}_{(h,q)}$
and $\stackrel{\hookleftarrow}{G}_{(h,q)}$, 
defined as follows: 
\bea 
\left(\Box_{\rm v} - \cM_{(h,q)}  \cM_{(h,q)}^\dagger \right) 
\stackrel{\hookrightarrow}{G}_{(h,q)} (z,z') 
&=& -  \mathbbm{1} \,\d^8(z-z') ~, \non \\
\left(\Box_{\rm v} - \cM_{(h,q)}^\dagger \cM_{(h,q)}   \right) 
\stackrel{\hookleftarrow}{G}_{(h,q)} (z,z') 
&=& -  \mathbbm{1} \,\d^8(z-z') ~.
\eea
The rationale for introducing the two different Green's functions 
lies in the fact  that the matrices $\cM_{(h,q)}$ and $\cM_{(h,q)}^\dagger $ 
do not commute in the deformed case \cite{KT},  
\bea
\big[ \cM_{(h,q)} \, , \cM^\dagger_{(h,q)} \big]\,  \S
&=& { h {\bar h} \over N} \, (q  - {1\over q} )^2\,
\Big\{\F \, {\rm tr} ( \F^\dagger \, \S)
- \F^\dagger \,{\rm tr} ( \F\, \S ) 
\Big\}~.
\label{MMdag}
\eea
In other words, 
using the terminology of linear algebra, 
$\cM_{(h,q)} $ is not a normal  operator
in the deformed case.
The two Green's functions coincide in the undeformed case, 
$\stackrel{\hookrightarrow}{G}_{(g,1)} 
= \stackrel{\hookleftarrow}{G}_{(g,1)} =\stackrel{{}} {G}_{(g,1)} $.
The propagators for the action (\ref{quad-prel}) are:
\bea
 {\rm i}  \, \langle  v (z)\, v^{\rm T} (z') \rangle &=& 
-   G_{(g,1)}(z,z') ~, \non \\
{\rm i}  \, \langle  \vf_1 (z)\,  \vf^\dagger_1 (z') \rangle &=& 
{1 \over 16} {\bar \cD}^2 \cD'^2 
G_{(g,1)}(z,z') ~, \quad
 \langle  \vf_1 (z)\, \vf^{\rm T}_1 (z') \rangle 
=0~. 
\eea
The propagators for the action (\ref{quad-prel2}) are:
\bea
{\rm i}  \, \langle  \vf_2 (z)\,  \vf^\dagger_2 (z') \rangle &=& 
{1 \over 16} {\bar \cD}^2 \cD'^2 
\stackrel{\hookleftarrow}{G}_{(h,q)} (z,z') ~, \non \\
{\rm i}  \, \langle  {\bar \vf}_3 (z)\,  \vf^{\rm T}_3 (z') \rangle &=& 
{1 \over 16} {\bar \cD}^2 \cD'^2 
\stackrel{\hookrightarrow}{G}_{(h,q)} (z,z') ~, \non \\
{\rm i}  \, \langle  \vf_2 (z)\,  \vf^{\rm T}_3 (z') \rangle &=&
\frac{1}{4} {\bar \cD}^2 \cM^\dagger_{(h,q)} 
\stackrel{\hookrightarrow}{G}_{(h,q)} (z,z') 
=\frac{1}{4} {\bar \cD}^2  
\stackrel{\hookleftarrow}{G}_{(h,q)} (z,z') \, \cM^\dagger_{(h,q)} ~, \\
{\rm i}  \, \langle {\bar \vf}_3 (z)\,  \vf^\dagger_2 (z') \rangle &=&
\frac{1}{4} \cD^2 \cM_{(h,q)} 
\stackrel{\hookleftarrow}{G}_{(h,q)} (z,z') 
=\frac{1}{4} \cD^2 
\stackrel{\hookrightarrow}{G}_{(h,q)} (z,z')\, \cM_{(h,q)} ~.
\non
\eea
In the above expressions for the propagators, 
all the fields are treated as adjoint column-vectors, 
in contrast to the Lie-algebraic notation used in 
the actions (\ref{quad-prel}) and (\ref{quad-prel2}). 
Due to the restrictions on the background superfields, 
eq. (\ref{back-con-1}), the Green's functions enjoy the 
following properties:
\bea
 \cD^2  \stackrel{\hookleftarrow}{G}_{(h,q)} (z,z')
= \cD'^2 
\stackrel{\hookleftarrow}{G}_{(h,q)} (z,z')~,
\qquad 
  {\bar \cD}^2  
\stackrel{\hookleftarrow}{G}_{(h,q)} (z,z') 
= {\bar \cD}'^2  
\stackrel{\hookleftarrow}{G}_{(h,q)} (z,z') ~,
\eea
and similarly for $\stackrel{\hookrightarrow}{G}_{(h,q)}$.

There are four supergraphs which contribute to the effective action 
at two loops -- three sunset graphs  constructed using 
the cubic vertices (\ref{hyper-vector}) and (\ref{hyper-3}), 
and one ``figure eight'' graph constructed using the quartic vertex (\ref{hyper-4}).
These supergraphs differ from the corresponding 
ones for the two-loop contribution to the effective action for $\cN=4$ SYM 
 in that the hypermultiplet propagators have deformed masses, 
whilst the sunset graph which originates from the cubic vertex $S^{(3)}_{\rm II}$ also 
has deformed group generators associated with the cubic vertices.

The contributions to the two-loop effective action from these supergraphs are 
(with traces in the adjoint representation):
\bea
 \Gamma_{\rm I} 
&=&  \frac{g^2}
{2^9}
\int \rd^8 z 
\, \rd^8 z' \, G^{ab}_{(g,1)}(z,z') \, \Big\{
\tr_{\rm Ad} \left(  T^a \,
 \bar{\cD}^2 \cD^2 
{\stackrel{\hookleftarrow}{G}_{(h,q)}} (z,z')
T^b \, \bar{\cD}'^2 \cD'^2  
{\stackrel{\hookleftarrow}{G}_{(h,q)} }
(z',z) \right)
\non
\\
&& \qquad \qquad +
\tr_{\rm Ad} \left(  T^a \,
 \cD^2 \bar{\cD}^2 
{\stackrel{\hookrightarrow}{G}_{(h,q)}} (z,z')
T^b \, \cD'^2 \bar{\cD}'^2   
{\stackrel{\hookrightarrow}{G}_{(h,q)}}
(z',z) \right) \Big\}~,
\non \\
 \Gamma_{\rm II} &=& - \, \frac{1}
{2^8}
\int \rd^8 z  
\, \rd^8 z' \, G^{ab}_{(g,1)}(z,z') \, 
\tr_{\rm Ad}
\Big(  
{T}_{(h,{1\over q})}^{\,a}
\bar{\cD}^2 \cD^2  
{ \stackrel{\hookleftarrow}{G}_{(h,q)}}
(z,z') 
{T}_{( \bar h ,q)}^{\,b}
\cD'^2 \bar{ \cD}'^2 
{\stackrel{\hookrightarrow}{G}_{(h,q)} }
(z',z) \Big),
\non
\\
 \Gamma_{\rm III} &=& - \, \frac{ g^2}{2^4} \int \rd^8 z
\, \rd^8 z' \, G^{ab}_{(g,1)}(z,z') \, \tr_{\rm Ad} \Big(  T^a \, \cM_{(h,q)}^{\dagger} \bar{\cD}^2 
{\stackrel{\hookrightarrow}{G}_{(h,q)} }(z,z') 
T^b \, \cM_{(h,q)} \, \cD'^2   
{\stackrel{\hookleftarrow}{G}_{(h,q)} }
(z',z) 
\Big),
\non  \\
 \Gamma_{\rm IV} &= &  
 \frac{g^2}{2^5} \int \rd^8 z \lim_{z' \rightarrow z} G^{ab}_{(g,1)}(z,z') \,\Big\{ 
\tr_{\rm Ad} \Big( T^a \, \bar{\cD}^2 \cD^2 
{\stackrel{\hookleftarrow}{G}_{(h,q)} }(z,z') 
T^b \Big)
\non \\
&& \qquad \qquad \qquad \qquad \quad \qquad \quad
+\tr_{\rm Ad} \Big( T^a \, \cD^2 \bar{\cD}^2  
{\stackrel{\hookrightarrow}{G}_{(h,q)} }(z,z') 
T^b \Big)
\Big\}~.
\label{I--IV}
 \eea

Before plunging into actual calculations,
it is instructive to give a  qualitative comparison of the quantum corrections 
(\ref{I--IV}) with those previously studied  for $\cN=4$ SYM \cite{KM3,KM4}.
In the absence of the deformation, i.e. in the case  $(h,q) = (g,1)$, 
all propagators are expressed via a single Green's function $G$
that, in the above notation, is 
$\stackrel{\hookrightarrow}{G}_{(g,1)} 
= \stackrel{\hookleftarrow}{G}_{(g,1)} =\stackrel{{}} {G}_{(g,1)} $, 
and the matrices ${T}_{(  h ,q)}$ in the expression for $\G_{\rm II} $ 
coincide with  the generators of $SU(N)$.
Then, the relative minus sign between the contributions $\Gamma_{\rm I} $ 
and $ \Gamma_{\rm II}$ allows them to be combined in the form
\be
\Gamma_{\rm I+II} = \frac{g^2}{2^8} \int \rd^8 z 
\,\rd^8 z' \, G^{ab}(z,z') \,
 \tr_{\rm Ad} \Big(  T^a \, \bar{\cD}^2 \cD^2 G (z,z') 
 T^b \,[ \bar{\cD}'^2 ,\cD'^2]  G  (z',z) \Big)~.
\ee
Using the properties of the superpropagators, 
this can be further manipulated to yield
\be
\Gamma_{\rm I+II} = \frac{g^2}{2^9} \int \rd^8 z 
\, \rd^8 z' \, G^{ab}(z,z') \, \tr_{\rm Ad} 
\Big(  T^a \, [\bar{\cD}^2 ,\cD^2] G (z,z') 
T^b \,[ \bar{\cD}'^2 ,\cD'^2]  G  (z',z) \Big)~.
\label{GammaI+II}
\ee
In conjunction with the identity
\be 
\frac{1}{16} [\cD^2 ,\bar{\cD}^2] 
= \frac{\ri}{4} \cD_{\ad} \cD^{\a \ad} \cD_{\a} -  \frac{\ri}{4} \cD_{\a} \cD^{\a \ad} \cD_{\ad} ~,
\label{id}
\ee
the above relation turns out to imply, in particular, that 
no effective K\"ahler potential is generated in $\cN=4$ SYM 
at two loops. The situation changes drastically in the $\b$-deformed theory. 

Let us first discuss the sunset diagrams $\G_{\rm I}$, $\G_{\rm II}$
and $\G_{\rm III}$ in (\ref{I--IV}). They all  involve a Green's function, 
$G^{\,ab}_{(g,1)}$, without spinor derivatives applied.
The latter proves to include, as a factor,  a (shifted) Grassmann delta-function
that can be used to eliminate one of the Grassmann integrals, 
say the one over $ \q'$. The two other Green's functions
are acted upon by some number $n\leq 4$ of spinor derivatives.
It can be shown that such a Green's function produces an overall factor 
of $\cW^{4-n}$, with $\cW$ standing for the spinor field strengths
$\cW_\a$ and ${\bar \cW}_\ad$ or their {\it vector} covariant derivatives.
If $n<4$, the corresponding supergraph does not generate any correction 
to the effective K\"ahler potential. For the supergraphs 
$\G_{\rm I}$ and $\G_{\rm II}$ in (\ref{I--IV}), we have $n=4$, 
and each of them gives rise to K\"ahler-like quantum corrections. 
In the case of $\cN=4$ SYM, the K\"ahler quantum corrections 
coming from $\G_{\rm I}$ and $\G_{\rm II}$ cancel each other, 
as a consequence of eqs. (\ref{GammaI+II}) and (\ref{id}).
In the deformed case, this cancellation does not take
place any more, and  two-loop corrections 
to the effective K\"ahler potential do occur.

As to the ``eight''  diagram  $\G_{\rm IV}$ in (\ref{I--IV}), 
it can be shown to produce an overall factor 
of $\cW^{4}$, similar to $\cN=4$ SYM,
and therefore no new effects occur in this sector.

So far the background superfields have been chosen to correspond 
to arbitrary directions in the Cartan subalgebra of $SU(N)$,
\bea
\F = 
{\rm diag} \,(\f^1, \dots ,  \f^N) ~, 
\quad \cW_\a = {\rm diag} \,(W^1_\a , \dots ,W^N_\a) ~, 
\quad \sum_{i=1}^{N} \f^i =  \sum_{i=1}^{N} W^i_\a =0~.
\eea
In what follows,  our consideration will be restricted  
to more special background scalar and vector superfields 
\be
\F =  \f \, H_0~, \ \ \ \ \  \qquad \cW_\a = W_\a \, H_0~,
\label{actual-background}
\ee
where $\f$  and $  W_\a $ are  singlet fields, and 
 $H_0$ has the form
\bea
H_0 =\frac{1}{\sqrt{N(N-1)} } {\rm diag}\,(N-1, -1,\cdots, -1)~.
\eea 
The  characteristic feature of this field configuration 
is that it leaves
the subgroup $U(1) \times SU(N-1) \subset SU(N)$ 
unbroken, where $U(1)$ is associated with $H_0$.
${}$For such background fields, 
the actual calculations turn out to
simplify drastically, and at the same time we are in a position to 
keep track of various effects induced by the deformation.
Among the simplifications which eq. (\ref{actual-background}) 
leads to, is that the fact that the mass matrices 
$\cM_{(h,q)}$ and $\cM_{(h,q)}^\dagger $ now  commute, 
\bea
\big[ \cM_{(h,q)} \, , \cM^\dagger_{(h,q)} \big]
&=&0~,
\label{MMdag2}
\eea
as can be seen  from (\ref{MMdag}).
As a consequence,  the Green's functions 
 $\stackrel{\hookrightarrow}{G}_{(h,q)}$
and $\stackrel{\hookleftarrow}{G}_{(h,q)}$
become identical, 
$\stackrel{\hookrightarrow}{G}_{(h,q)} =\stackrel{\hookleftarrow}{G}_{(h,q)}
\equiv {G}_{(h,q)}$. For the background chosen, one can also check 
the validity of 
the identity  $(h\f)^{-1} \cM_{(h,q)}^{\rm T} = - ({\bar h} {\bar \f})^{-1} 
\cM^\dagger_{(h,q)} \,$, which leads to the important symmetry property 
\bea
\Big(G_{(h,q)}(z,z') \Big)^{\rm T} = G_{(h,q)}(z',z) ~.
\eea
The two-loop 
contributions to the effective action
become
\bea
 \Gamma_{\rm I+II} 
&=&  \frac{1}
{2^9}
\int \rd^8 z 
\, \rd^8 z' \, G^{ab}_{(g,1)}(z,z') \, \Big\{ g^2 \,
\tr_{\rm Ad} \Big(  T^a \,
[ \bar{\cD}^2 , \cD^2 ] 
{G}_{(h,q)} (z,z')
T^b \, 
[ \bar{\cD}'^2 , \cD'^2 ] 
{G}_{(h,q)} 
(z',z) \Big)
\non
\\
&& \qquad \qquad +\,
2\, g^2 \, \tr_{\rm Ad} \Big(  T^a \,
  \bar{\cD}^2 \cD^2
{G}_{(h,q)} (z,z')
T^b \, \cD'^2 \bar{\cD}'^2   
G_{(h,q)}   
(z',z) 
\Big)
\non \\
&& \qquad \qquad 
-\,2\,\tr_{\rm Ad}
\Big(  
{T}_{(h,{1\over q})}^{\,a}
\bar{\cD}^2 \cD^2  
{G}_{(h,q)}
(z,z') 
{T}_{( \bar h ,q)}^{\,b}
\cD'^2 \bar{ \cD}'^2 
{G}_{(h,q)} 
(z',z) \Big) \Big\}~, \label{two-loop} \\
 \Gamma_{\rm III} &=& - \, \frac{ g^2}{2^4} \int \rd^8 z
\, \rd^8 z' \, G^{ab}_{(g,1)}(z,z') \, \tr_{\rm Ad} \Big(  T^a \, \cM_{(h,q)}^{\dagger} \bar{\cD}^2 
{G}_{(h,q)} (z,z') 
T^b \, \cM_{(h,q)} \, \cD'^2   
{G}_{(h,q)} 
(z',z) 
\Big),
\non  \\
 \Gamma_{\rm IV} &= &  
 \frac{g^2}{2^4} \int \rd^8 z \lim_{z' \rightarrow z} G^{ab}_{(g,1)}(z,z') \,
\tr_{\rm Ad} \Big( T^a \, \bar{\cD}^2 \cD^2  
{G}_{(h,q)} (z,z') 
T^b \Big)~.
\non
\eea
In the expression for $  \Gamma_{\rm I+II} $,
it is only the contributions in the second and third lines
which generate the effective K\"ahler potential.

\sect{Decomposition into \mbox{$\bm{U(1)}$} Green's functions}
As in the undeformed case \cite{KM3,KM4}, 
in the presence of  the special background (\ref{actual-background}), 
the expressions (\ref{two-loop}) for the two-loop contributions to the effective action 
decompose into a set of terms involving only $U(1)$ Green's functions, as outlined below.

The generic group theoretic structure of 
$\Gamma_{\rm I+II}$ and $\Gamma_{\rm III}$ is
\be \Gamma = \kappa \int \rd^8 z \int \rd^8 z' \, G^{ab} \, \tr_{\rm Ad} \Big({T}_{(h,{1\over q})}^{\,a} \, \hat{ G }_{(h,q)} \, 
{T}_{(\bar{h},q)}^{\,b}\, \check{G  }^{\, \prime}_{(h,q)}  \Big)~,
\label{genericGamma}
\ee
where $G^{ab}$ is an undeformed Green's function, 
$ \hat{ G }_{(h,q)} $ and $\check{ {G } }_{(h,q)} $ denote spinor derivatives of 
the deformed  Green's function $G_{(h,q)}$ (multiplied by mass matrices 
in the case of $\Gamma_{\rm III}),$ and unprimed Green's functions have 
argument $(z, z'),$ primed Green's functions have argument $(z',z).$ 
Contributions with undeformed group generators are obtained by setting 
$h=g$ and $q = 1$ in ${T}_{(h,{1\over q})}^{\,a}$ and ${T}_{(\bar{h},q)}^{\,b}.$

Relative to the standard Cartan basis\footnote{The index $I$ takes 
the values $0, 1, 2, \cdots, N-2,$ while the index $i$ takes the values $0,
 \underline{i}$ with $\underline{i} = 1, 2, \cdots, N-1.$} $(H_I, E_{ij})$ 
for the Lie algebra of $SU(N)$, 
which is explicitly given  in Appendix B, 
the Green's functions have the decomposition
\be
G_{(h,q)}^{ab}=  \left( \begin{array}{cc} 
G_{(h,q)}^{IJ} & 0 \\ 0 & G_{(h,q)}^{ij,kl} \end{array} \right)~.
\ee
When the background corresponds to an arbitrary direction 
in the Cartan subalgebra  of $SU(N)$ (i.e. prior to the choice of the special
 background (\ref{actual-background})), the structure of the 
deformed mass matrix 
(\ref{massop}) is such that only the diagonal entries $G^{ij,ji}$ are nonzero, 
whereas $G^{IJ}$ is not diagonal. The expression (\ref{genericGamma}) 
therefore decomposes as
\bea
\Gamma &= & \kappa \,  |h|^2  \int \rd^8 z \int \rd^8 z'  
\left\{  ( G^{ij,ji} \, \hat{G}^{IJ}_{(h,q)}  \, \check{G}^{\, \prime \,ij,ji}_{(h,q)}  + G^{ij,ji} \, \hat{G}^{ji,ij}_{(h,q)}  \, 
\check{G}^{\, \prime \,IJ}_{(h,q)}  + G^{JI} \, \hat{G}^{ij,ji}_{(h,q)}  \, \check{G}^{\, \prime \, ij,ji}_{(h,q)}  ) \right. 
\nonumber \\ 
& & \times  \left( q^{-1}  (H_I)_{jj} - q (H_I)_{ii} \right) \, \left( q (H_J)_{jj} - q^{-1}
  (H_J)_{ii} \right) \nonumber \\
&+& G^{ij,ji} \left( \hat{G}^{li,il}_{(h,q)}  \check{G}^{\, \prime \, lj,jl}_{(h,q)}  
+ \hat{G}^{jl,lj}_{(h,q)}  \check{G}^{\, \prime \, il,li}_{(h,q)}  \right) \nonumber \\
&+& \left.  | q - q^{-1}|^2 \, G^{IJ} \hat{G}^{KL}_{(h,q)}  \check{G}^{\, \prime \, NM}_{(h,q)}  \, 
\tr_{F} (H_I H_K H_M) \, \tr_F (H_J H_L H_N) \right\} \nonumber \\
& \equiv & \Gamma_A + \Gamma_B + \Gamma_C~.
\label{GammaABC}
\eea

Specializing to the background (\ref{actual-background}) which breaks 
the $SU(N)$ gauge group to $SU(N-1)\times U(1),$ $G_{(h,q)}^{IJ}$ 
also becomes diagonal, and so $G_{(h,q)}^{ab}$ decomposes into  
a set of $U(1)$ Green's functions on the diagonal. With the notation 
that ${\bf G}^{(e)}$ denotes a deformed $U(1)$ Green's function 
with charge $e$, 
\be
e=  \sqrt{\frac{N}{N-1} }~,
\ee
 relative to the generator $H_0$ of the Cartan subalgebra,
\be
G_{(h,q)}^{IJ} =  {\rm diag }( \tilde{{\bf G}}^{(0)}, {\bf G}^{(0)}, {\bf G}^{(0)}, \cdots , {\bf G}^{(0)}) 
\ee
and 
\be 
G_{(h,q)}^{0 \underline{i}, \underline{i} 0} 
= {\bf G}^{(e)}, \quad G_{(h,q)}^{ \underline{i} 0, 0 \underline{i} } = {\bf G}^{(-e)}, \quad 
G_{(h,q)}^{ \underline{i} \underline{j}, \underline{j} \underline{i}} = {\bf G}^{(0)}~.
\ee 
The eigenvalues of the mass matrix (\ref{massop})
 associated with the deformed $U(1)$ Green's functions are:
\bea
\tilde{{\bf G}}^{(0)} : \tilde{{\bf M}}^{(0)}  &=&  h (q - q^{-1})\, 
\frac{(N-2)}{\sqrt{N (N-1)}}  \, \phi ~,  \nonumber \\
{\bf G}^{(0)} : {\bf M}^{(0)}  &=&  - h (q - q^{-1}) \,  \frac{1}{\sqrt{N (N-1)}}  \, \phi 
~, \nonumber \\
{\bf G}^{( e)}  : {\bf M}^{( e)}  &=& -h (q + (N-1) q^{-1}) \,  \frac{1}{\sqrt{N (N-1)}}  \, \phi 
~, \nonumber \\
{\bf G}^{(-e)}  : {\bf M}^{(-e)}   &=& h ((N-1)q + q^{-1}) \,  \frac{1}{\sqrt{N (N-1)}}  \, \phi~ .
\label{massentries} 
\eea
It is worth reiterating that the deformed $U(1)$ Green's functions occur 
only in the hypermultiplet propagators - the vector and chiral scalar 
Green's functions and corresponding masses are obtained by setting 
$h=g$ and $q = 1.$ An undeformed $U(1)$ Green's function of charge $e$ 
will be denoted $G^{(e)}$
i.e. $G^{(e)} = {\bf G}^{(e)}|_{h=g,q=1}.$ Note that $\tilde{{\bf G}}^{(0)}$ 
and $ {\bf G}^{(0)} $  become the same massless Green's function $G^{(0)} $ 
when the deformation vanishes.

In the special  background (\ref{actual-background}), the expressions for 
$ \Gamma_A,$ $ \Gamma_B$ and $ \Gamma_C$ in (\ref{GammaABC}) 
decompose into contributions involving only $U(1)$ Green's functions:
\bea
\Gamma_A &=& \kappa  |h|^2  \int \rd^8 z \int \rd^8 z'  
\left\{ N \, \Big(1 - |q - q^{-1}|^2 \frac{(N-1)}{N^2} \Big) \,  (G^{(-e)} 
\hat{\tilde{{\bf G}}}^{(0)} \check{{\bf G}}'^{(-e)} \right.  \nonumber \\ 
&+&  G^{(-e)} \hat{{\bf G}}^{(e)} \check{\tilde{{\bf G}}}'^{(0)} 
+G^{(0)} \hat{{\bf G}}^{(-e)} \check{{\bf G}}'^{(-e)} )
 +  (N-2) \,  (G^{(-e)} \hat{{\bf G}}^{(0)} \check{{\bf G}}'^{(-e)}  \nonumber \\ 
&+& G^{(-e)} \hat{{\bf G}}^{(e)} \check{{\bf G}}'^{(0)} + G^{(0)} 
\hat{{\bf G}}^{(-e)} \check{{\bf G}}'^{(-e)} )
+ (e \leftrightarrow -e) \nonumber \\ 
&+&  \frac{(N-2)}{N} |q - q^{-1}|^2 \, ( G^{(0)} \hat{{\bf G}}^{(0)} \check{{\bf G}}'^{(0)} 
+G^{(0)} \hat{\tilde{{\bf G}}}^{(0)} \check{{\bf G}}'^{(0)} + G^{(0)} \hat{{\bf G}}^{(0)} 
\check{\tilde{{\bf G}}}'^{(0)} ) \nonumber \\ 
& + & \left. 6 (N-1) (N-2) \Big( 1 - \frac{|q - q^{-1}|^2 }{2 (N-1)} \Big) \, 
G^{(0)} \hat{{\bf G}}^{(0)} \check{{\bf G}}'^{(0)} \right\}~,
\eea
\bea
\Gamma_B &=&  \kappa |h|^2 \, (N-1)(N-2) \, \int \rd^8 z \int \rd^8 z'  \left\{
(G^{(e)} \hat{{\bf G}}^{(-e)} \check{{\bf G}}'^{(0)} + G^{(e)} \hat{{\bf G}}^{(0)} 
\check{{\bf G}}'^{(e)}   \right. \nonumber \\ &+ &  G^{(0)} \hat{{\bf G}}^{(e)} 
\check{{\bf G}}'^{(e)} )+ \left.  (e \leftrightarrow -e) + 2  (N-3) \, 
G^{(0)} \hat{{\bf G}}^{(0)} \check{{\bf G}}'^{(0)} \right\} ~,
\eea
\bea
\Gamma_C &=& \kappa |h|^2 |q - q^{-1}|^2 \, \frac{(N-2)}{N (N-1)} 
\int \rd^8 z \int \rd^8 z'  \, G^{(0)}  \left\{ (N-2) \,  \hat{\tilde{{\bf G}}}^{(0)} 
\check{\tilde{{\bf G}}}'^{(0)}    \right. \nonumber \\ &+ & \left. 
\hat{\tilde{{\bf G}}}^{(0)} \check{{\bf G}}'^{(0)} +  \hat{{\bf G}}^{(0)}
 \check{\tilde{{\bf G}}}'^{(0)} + (N^2-3N+1) \,   \hat{{\bf G}}^{(0)} 
\check{{\bf G}}'^{(0)} \right\}~.
\eea

Using the above  results,  $\Gamma_{\rm I + II},$  $\Gamma_{\rm III}$ and
 $\Gamma_{\rm IV}$ can be expressed in terms of $U(1)$ Green's functions. 
Adopting the specific notation 
$ \hat{G} = \bar{\cD}^2 \cD^2 G(z,z'), $ $\check{G}^{\, \prime} 
= \cD'^2 \bar{\cD}'^2 G(z',z), $ for $\Gamma_{\rm I + II}$
one gets 
\bea
\Gamma_{\rm I + II} &=&
\Gamma_{\rm I + II}^{(A)} 
+\Gamma_{\rm I + II}^{(B)}~,
\eea
where
\bea
 \Gamma_{\rm I + II}^{(A)} &=&
 \frac{g^2}{2^9}Ê \int \rd^8 z \int \rd^8 z' \, N \,
\Big\{ (N-1) \,Ê G^{(0)} [\bar{\cD}^2, \cD^2] {\bf G}^{(e)} [\bar{\cD}'^2, \cD'^2] 
{\bf G}^{ \prime \, (e)} 
\nonumber \\
& + & (N-2) G^{(e)}Ê\Big(
 [\bar{\cD}^2, \cD^2] {\bf G}^{(0)} [\bar{\cD}'^2, \cD'^2] 
{\bf G}^{ \prime \, (e)}
+[\bar{\cD}^2, \cD^2] {\bf G}^{(-e)} [\bar{\cD}'^2, \cD'^2] 
{\bf G}^{ \prime \, (0)} \Big) \nonumber \\
&+& G^{(e)} \Big( [\bar{\cD}^2, \cD^2]\tilde{{\bf G}}^{(0)}[\bar{\cD}'^2, \cD'^2] 
{\bf G}^{ \prime \, (e)}
+ [\bar{\cD}^2, \cD^2]{\bf G}^{(-e)}[\bar{\cD}'^2, \cD'^2] 
\tilde{{\bf G}}^{ \prime \, (0)}\Big)
+ (e \leftrightarrow -e) \nonumber \\
&+ &  2 (N-1) (N-2) G^{(0)}Ê [\bar{\cD}^2, \cD^2] {\bf G}^{(0)} 
[\bar{\cD}'^2, \cD'^2] {\bf G}^{ \prime \, (0)} \Big\} 
~,
\label{G:I+II-A1}
\eea
and 
\bea
\Gamma_{\rm I + II}^{(B)}
&=&
\frac{1}{2^8} \,Ê \int \rd^8 z \int \rd^8 z'Ê \, \left\{Ê N (N-1) 
\Big(g^2 - |h|^2 \big(1 - \frac{|q-q^{-1}|^2}{N^2}\big) \Big)Ê \, G^{(0)} 
\hat{{\bf G}}^{(e)} \check{{\bf G}}^{\prime \, (e)} \right. \nonumber \\
&+ &ÊÊ N(N-2)(g^2 - |h|^2) G^{(e)} \big( \hat{{\bf G}}^{(0)} 
\check{{\bf G}}^{\prime \, (e)}ÊÊ +  \hat{{\bf G}}^{(-e)}
 \check{{\bf G}}^{\prime \, (0)} \big) \nonumber \\
&+& N \Big( g^2 - |h|^2 \big(1 -Ê |q-q^{-1}|^2 \frac{(N-1) }{N^2} \big) \Big) \, 
G^{(e)} \big(  \hat{\tilde{{\bf G}}}^{(0)} \check{{\bf G}}^{\prime \, (e)} 
+ \hat{{\bf G}}^{(-e)} \check{\tilde{{\bf G}}}^{\prime \, (0)} \big)
\nonumber \\
&+& (e \leftrightarrow -e) \nonumber \\
& + &Ê 2 N (N-1)(N-2) \Big( g^2 - |h|^2 \big(1 -Ê |q - q^{-1}|^2 
\frac{(2N-1)}{2N(N-1)^2} \big) \Big)
\,Ê G^{(0)} \hat{{\bf G}}^{(0)} \check{{\bf G}}^{\prime \, (0)}Ê \nonumber \\
&-& |h|^2 \, |q - q^{-1}|^2 \frac{(N-2)}{(N-1)} G^{(0)}Ê \, \Big(  
\hat{\tilde{{\bf G}}}^{(0)} \check{{\bf G}}^{\prime \, (0)} +Ê \hat{{\bf G}}^{(0)} 
\check{\tilde{{\bf G}}}^{\prime \, (0)} \nonumber \\
&+& \left. \frac{(N-2)}{N}  \hat{\tilde{{\bf G}}}^{(0)} 
\check{\tilde{{\bf G}}}^{\prime \, (0)} \Big) \right\} .
\eea
If one takes into account the one-loop finiteness condition, 
eq. (\ref{finite2}), then the expression for $\Gamma_{\rm I + II}^{(B)}$
simplifies considerably 
\bea
\Gamma_{\rm I + II}^{(B)}
&=&-
\frac{1}{2^8} \,Ê 
\frac{N-2}{N} |h|^2 
|q - q^{-1}|^2 
\int \rd^8 z \int \rd^8 z'Ê \, \left\{Ê 
G^{(e)} 
\big( \hat{{\bf G}}^{(0)} 
\check{{\bf G}}^{\prime \, (e)}ÊÊ + 
\hat{{\bf G}}^{(-e)}
 \check{{\bf G}}^{\prime \, (0)} \big) 
\right.
 \nonumber \\
&&- 
G^{(e)} \big( \hat{\tilde{{\bf G}}}^{(0)} \check{{\bf G}}^{\prime \, (e)} 
+ \hat{{\bf G}}^{(-e)} \check{\tilde{{\bf G}}}^{\prime \, (0)} \big)
+(e \leftrightarrow -e) \nonumber \\
&&+ 
\frac{N}{N-1}\, 
G^{(0)}Ê 
\Big(
\hat{\tilde{{\bf G}}}^{(0)} \check{{\bf G}}^{\prime \, (0)} 
+Ê \hat{{\bf G}}^{(0)} 
\check{\tilde{{\bf G}}}^{\prime \, (0)} 
\nonumber \\
&&\qquad \qquad \qquad 
+\frac{(N-2)}{N} 
\hat{\tilde{{\bf G}}}^{(0)} 
\check{\tilde{{\bf G}}}^{\prime \, (0)} 
- \frac{3N-2}{N}\,Ê
  \hat{{\bf G}}^{(0)} \check{{\bf G}}^{\prime \, (0)}Ê
\Big) \Big\}~ .
\label{G:I+II-B1}
\eea

With the notation
$\hat{G} = \bar{\cD}^2 G(z,z'),$ $ \check{G}^{\, \prime} = \cD'^2 G(z',z), $
\bea
\Gamma_{\rm III} &=& - \frac{g^2}{2^4}  \int \rd^8 z \int \rd^8 z' \, 
N \,\left\{ (N-1) \, {\bf M}^{(e) \dagger} {\bf M}^{(e)} G^{(0)} \hat{{\bf G}}^{(e)} 
\check{{\bf G}}^{ \prime \, (e)} \right. \nonumber \\ 
& + & (N-2) {\bf M}^{(0) \dagger} {\bf M}^{(e)} G^{(e)} \hat{{\bf G}}^{(0)} 
\check{{\bf G}}^{ \prime \, (e)} 
+ (N-2) \, {\bf M}^{(-e) \dagger} {\bf M}^{(0)} G^{(e)} \hat{{\bf G}}^{(-e)}
 \check{{\bf G}}^{ \prime \, (0)} \nonumber \\
& + & \tilde{{\bf M}}^{(0) \dagger} {\bf M}^{(e)} G^{(e)} \hat{\tilde{{\bf G}}}^{(0)} 
\check{{\bf G}}^{ \prime \, (e)} 
+ {\bf M}^{(-e) \dagger} \tilde{{\bf M}}^{(0)} G^{(e)} \hat{{\bf G}}^{(-e)}
 \check{\tilde{{\bf G}}}^{ \prime \, (0)} 
 + (e \leftrightarrow -e) \nonumber \\
 &+ & \left. 2 (N-1) (N-2) {\bf M}^{(0) \dagger}  {\bf M}^{(0) } G^{(0)} 
\hat{{\bf G}}^{(0)} \check{{\bf G}}^{ \prime \, (0)} \right\}~.
\label{G-III}
\eea

${}$Finally, the group theory involved in evaluating $\Gamma_{\rm IV}$ 
is relatively straightforward, as it does not involve deformed generators.  
With the notation $ \hat{G} = \bar{\cD}^2 \cD^2G(z,z'), $
\bea
\Gamma_{\rm IV} &=&  \frac{g^2}{2^4}  \int \rd^8 z \lim_{z' \rightarrow z} 
\left\{ N (N-1)\, G^{(0)} \Big( \hat{{\bf G}}^{(e)} + \hat{{\bf G}}^{(-e)} 
+ 2 (N-2) \hat{{\bf G}}^{(0)} \Big) \right.  \nonumber \\
& + & N G^{(e)} \Big( \hat{\tilde{{\bf G}}}^{(0)}  
+ (N-2) \hat{{\bf G}}^{(0)} + (N-1) \hat{{\bf G}}^{(e)}  \Big) \nonumber \\
& + & \left. N G^{(-e)} \Big( \hat{\tilde{{\bf G}}}^{(0)}  
+ (N-2) \hat{{\bf G}}^{(0)} + (N-1) \hat{{\bf G}}^{(-e)}  \Big) \right\}~ .
\label{G-IV}
\eea

Let us list all the masses appearing in the theory: 
\bea
m_1^2 &=&g^2 \frac{N}{N-1} \f \bar \f ~,\non \\
m_2^2 &=& |h|^2
\frac{| q-q^{-1}|^2 
}{N(N-1)} \f \bar \f  ~, \non \\
m_3^2 &=& |h|^2
\frac{| q-q^{-1}|^2
(N-2)^2}{N(N-1)} \f \bar \f  ~, \non \\
m_4^2 &=& |h|^2  \frac{| q (N-1)+q^{-1}|^2}{N(N-1)} \f \bar \f  ~.
\label{m1--4}
\eea 
Here $m_1$ is the undeformed mass. 
It corresponds to the Green's function $G^{(e)}$.
The masses $m_2$, $m_3$ and $m_4$, which involve
the deformation parameter,  correspond to the Green's functions 
${\bf G}^{(0)}$,
$\tilde{{\bf G}}^{(0)}$ and ${\bf G}^{(e)}$, respectively.  

Looking at the structure of the specific supergraphs contributing 
to $\G_{\rm I +II}$ and $\G_{\rm III}$, one can see that there occur
only two different mass assignments  with all non-vanishing masses: 
(i) $m_1$, $m_2$ and
$m_4$; (ii) $m_1$, $m_3$ and $m_4$. In these cases  
\bea
\D(m_1^2, m_2^2, m_4^2) &=&|h|^4\,
\frac{  | q-q^{-1}|^2}{(N-1)^2} \,
\Big\{ 4
 - \frac{(N-1)^2 +4}{N^2} | q-q^{-1}|^2\Big\} \, 
(\f \bar \f )^2~,  \\
\D(m_1^2, m_3^2, m_4^2) &=&|h|^4\,
\frac{  | q-q^{-1}|^2 (N-2)^2}{(N-1)^2} \,
\Big\{ 4
 - \frac{(N-1)^2 +4}{N^2} | q-q^{-1}|^2\Big\} \, 
(\f \bar \f )^2~. \non 
\eea
${}$For $q\neq \pm 1$ and large finite $N$, 
both cases are characterized by  $\D>0$, and therefore 
one has to deal with  two-loop integrals 
$ I(\n_1 ,\n_2 , \n_3 ; x,y,z)$ satisfying this condition. 
Such integrals are studied in detail in
Appendix \ref{two-loop-int}. Only in the limit
$q\to \pm1$ is  the BPS condition $\D=0$  restored.
It should be pointed out that there supergraphs 
with two non-vanishing masses also occur, and in these cases
 one has to deal with  two-loop integrals 
$ I(\n_1 ,\n_2 , \n_3 ; 0,y,z)$.

\sect{Covariant superpropagators 
and dimensional reduction}
In the remainder of this paper, we 
evaluate two-loop
quantum corrections of the form:
\bea
\int{\rm d}^8 z \,K(\f, \bar \f )
+ c\int{\rm d}^8 z \,\frac{ W^2 {\bar W}^2}{\f^2  {\bar \f}^2 } ~.
\label{gluestructure}
\eea
This can be achieved by considering a simplest choice 
of  constant background chiral superfields:
\be
\f ={\rm const}~, \qquad W_\a ={\rm const}~.
\label{glueball}
\ee

The first term in (\ref{gluestructure}) corresponds to the effective K\"ahler 
potential, and its origin is solely due to the $\b$-deformation. 
Since the background superfields correspond to the very special 
direction in the Cartan subalgebra of $SU(N)$, eq. (\ref{actual-background}),
the $\cN=1$ 
superconformal invariance requires $K(\f, \bar \f ) \propto \f \bar \f $.
For a more general choice of background superfields. 
the effective K\"ahler potential is expected to receive more complicated corrections 
of the form 
\be
 \sum_{i<j}
|\f^i -\f^j |^2 
\,F \Big( \Big| \frac{   (   \f^i -   \f^j    )^2}{  ( q\, \f^i -  q^{-1} \, \f^j )
( q^{-1}\, \f^i -  q \, \f^j )    } 
\Big| \Big)~,
\ee
which are compatible with superconformal invariance. 

The second  term in (\ref{gluestructure}) is known to be superconformally 
invariant, and it generates $F^4$ terms at the component level. 
Such quantum corrections are of some interest in the context 
of supergravity--gauge theory duality in the description of D-brane interactions, 
see e.g. \cite{BKT,KT} and references therein.

Expressions for $U(1)$ Green's functions of the type given in section 4 
are known in closed form \cite{KM,KM2} in the case when  the background vector 
multiplet obeys the constraint $D_\a W_\b = {\rm const}$, 
which is weaker than (\ref{glueball}).
Under the constraints (\ref{glueball}),
the Green's function of charge $e$ and mass $m$ is  
\be
G^{(e)}(z,z') = 
{\rm i} \int\limits_0^\infty {\rm d}s \, K^{(e)}(z,z',{\rm i}s) ~,
\label{proper-time-repr}
\ee
where the heat kernel has the form
\bea
K^{(e)}(z,z',{\rm i}s) &=& \frac{\rm i}{(4 \pi {\rm i}s)^2} \, 
{\rm e}^{ {\rm i}\r^2 /4s  -{\rm i}(m^2 -{\rm i}0 ) s} \, 
\d^2(\z  -{\rm i} s \,\cW) \,
\d^2({\bar \z}  + {\rm  i}s \,{\bar \cW} ) 
\, I(z,z')~, \non \\
&\equiv& K_0 (\r , {\rm i}s | m^2)  \,
\d^2(\z  -{\rm i} s \,\cW) \,
\d^2({\bar \z}  + {\rm  i}s \,{\bar \cW} ) 
\, I(z,z')~,
\eea
and we have introduced the supersymmetric two-point functions
\be
\r^a = (x-x')^a - {\rm i} (\q-\q') \s^a {\bar \q}' 
+ {\rm i} \q' \s^a ( {\bar \q} - {\bar \q}') ~, \quad
\z^\a = (\q - \q')^\a ~, \quad
{\bar \z}_\ad =({\bar \q} -{\bar \q}' )_\ad ~, 
\label{two-point}
\ee
the field strengths $\cW_\a = e W_\a$ and 
${\bar \cW}_\ad = e {\bar W}_\ad$, 
and the parallel displacement propagator $I(z,z')$
(see \cite{KM} for more details)
which  is  completely 
specified by the properties:
\bea
I(z',z) \, \cD_{\a \ad} I(z,z') &=&
-{\rm i} ( \z_{\a} \bar{\cW}_{\ad} 
+ \cW_\a \, \bar{\z}_{\ad}   ) ~, \non \\
I(z',z) \,  \cD_{\a} I(z,z') &=&
- \frac{{\rm i}}{2} \, \r_{\a \ad} 
\bar{\cW}^{\ad} 
+ \frac13 ( \z_{\a} \bar{\z} \bar{\cW}
+ \bar{\z}^2  \cW_{\a})~, \\
I(z',z) \,  {\bar \cD}_{\ad} I(z,z') &=&
- \frac{{\rm i}}{2} \, \r_{\a \ad} \cW^{\a} 
- \frac13 ( {\bar \z}_\ad \z \cW  + \z^2 {\bar \cW}_\ad)~. \non
\eea
The chiral kernel becomes
\bea
K^{(e)}_+(z,z', {\rm i}s) &=& -\frac{1 }{ 4} {\bar \cD}^2 K^{(e)}(z,z'|s) \non \\
&=&
K_0 (\r , {\rm i}s | m^2)  \,
\d^2(\z  -{\rm i} s \,\cW) \,
{\rm e}^{ \frac{{\rm i}}{6} s\,\cW^2 \,({\bar \z} +{\rm i}s\, {\bar \cW})^2}
\, I(z,z')~. 
\label{chiral-glueball}
\eea
Next, we obtain
\bea
&-&\frac{1 }{ 4} \cD^2 K^{(e)}_+(z,z', {\rm i}s) =
K_0 (\r , {\rm i}s | m^2)  \,
\exp \Big\{ 
\frac{1}{2s} 
\r^a (\z  -{\rm i} s \,\cW) \s_a ({\bar \z} +{\rm i}s\, {\bar \cW}) \Big\} \\
&& \qquad  \times  \exp \Big\{ \frac{\rm i }{ 2s } \z^2 {\bar \z}^2 
+\frac{2}{3} (\z \cW \,{\bar \z}^2 - \z^2 \,{\bar \z} {\bar \cW})
+{\rm i}s  \z \cW \,  {\bar \z} {\bar \cW}
-\frac{\rm i}{6}s^3 \cW^2 \,{\bar \cW}^2 \Big\} I(z,z')~.
\non
\eea
In the Grassmann coincidence limit, this reduces to 
\bea
\frac{1 }{ 16} \cD^2  {\bar \cD}^2
K^{(e)}(z,z', {\rm i}s)\Big|_{\z =0}
 &=&
K_0 (\r , {\rm i}s | m^2)  \,
\exp \Big\{ 
\frac{s}{2} \, 
\r^a \cW \s_a {\bar \cW}
-\frac{\rm i}{6} s^3\, \cW^2 \,{\bar \cW}^2
 \Big\} I~.
\label{Gcl1}
\eea

The antichiral kernel becomes
\bea
K^{(e)}_-(z,z', {\rm i}s) &=& -\frac{1 }{ 4} \cD^2 K^{(e)}(z,z'|s) \non \\
&=&
K_0 (\r , {\rm i}s | m^2)  \,
\d^2({\bar \z}  +{\rm i} s \,{\bar \cW}) \,
{\rm e}^{ \frac{{\rm i}}{6} s\,{\bar \cW}^2 \,(\z -{\rm i}s\, \cW)^2}
\, I(z,z')~,
\label{antichiral-glueball}
\eea
and so
\bea
&-&\frac{1 }{ 4} {\bar \cD}^2 K^{(e)}_-(z,z', {\rm i}s) =
K_0 (\r , {\rm i}s | m^2)  \,
\exp \Big\{ 
-\frac{1}{2s} 
\r^a (\z  -{\rm i} s \,\cW) \s_a ({\bar \z} +{\rm i}s\, {\bar \cW}) \Big\} \\
&& \qquad  \times  \exp \Big\{ \frac{\rm i }{ 2s } \z^2 {\bar \z}^2 
+\frac{2}{3} (\z \cW \,{\bar \z}^2 - \z^2 \,{\bar \z} {\bar \cW})
+{\rm i}s  \z \cW \,  {\bar \z} {\bar \cW}
-\frac{\rm i}{6}s^3 \cW^2 \,{\bar \cW}^2 \Big\} I(z,z')~.
\non
\eea
In the Grassmann coincidence limit, this reduces to 
\bea
\frac{1 }{ 16}   {\bar \cD}^2 \cD^2  
K^{(e)}(z,z', {\rm i}s)\Big|_{\z =0}
 &=&
K_0 (\r , {\rm i}s | m^2)  \,
\exp \Big\{
-\frac{s}{2} \, 
\r^a \cW \s_a {\bar \cW}
-\frac{\rm i}{6} s^3\, \cW^2 \,{\bar \cW}^2
 \Big\} I~.~~~~
 \label{Gcl2}
\eea

Using the above results, one readily obtains
\bea
&&\frac{1 }{ 16}  \big[ {\bar \cD}^2,  \cD^2  \big]
K^{(e)}(z,z',{\rm i}s) = -\frac{1}{s} K_0 (\r , {\rm i}s | m^2)  \,
\r^a (\z  -{\rm i} s \,\cW) \s_a ({\bar \z} +{\rm i}s\, {\bar \cW}) \\
&& \qquad  \times  \exp \Big\{ \frac{\rm i }{ 2s } \z^2 {\bar \z}^2 
+\frac{2}{3} (\z \cW \,{\bar \z}^2 - \z^2 \,{\bar \z} {\bar \cW})
+{\rm i}s  \z \cW \,  {\bar \z} {\bar \cW}
-\frac{\rm i}{6}s^3 \cW^2 \,{\bar \cW}^2 \Big\} I(z,z')~.
\non
\eea

The supersymmetric regularization by dimensional reduction 
\cite{Siegel} is implemented as follows:
\bea
 \frac{\rm i}{(4 \pi {\rm i}s)^2} \, 
{\rm e}^{ {\rm i}\r^2 /4s  -{\rm i}(m^2 -{\rm i}0 ) s}~ ~\longrightarrow ~~
\frac{\rm i}{(4 \pi {\rm i}s)^{d/2}} \, 
{\rm e}^{ {\rm i}\r^2 /4s  -{\rm i}(m^2 -{\rm i}0 ) s} 
\equiv \cK_0 (\r , {\rm i}s | m^2) ~.
\label{d-kernel}
\eea
Here and in what follows, 
for the free heat kernel in $d$ dimensions, we use 
the  notation, $\cK_0 (\r , {\rm i}s | m^2)$. 
The Green's function generated by $\cK_0 (\r , {\rm i}s | m^2)$
is denoted $\cG_0 (\r  | m^2)$. The free heat kernel in $d=4$ 
is denoted $K_0 (\r , {\rm i}s | m^2)$. 

\sect{Evaluation of two-loop quantum corrections}

This section is devoted to the calculation of 
the two-loop quantum corrections of the form (\ref{gluestructure}) 
that are generated by 
the supergraphs
listed in section 4.

\subsection{K\"ahler potential}
The two-loop quantum corrections to the K\"ahler potential 
are generated only by the functional  $ \Gamma_{\rm I + II}^{(B)} $,
eq. (\ref{G:I+II-B1}). To evaluate them, one can set 
$W_\a = {\bar W}_\ad =0$ in the propagators described 
in the previous section.
One thus obtains
\bea
K(\f, \bar \f )&=&- 4
\frac{N-2}{N}
(\m^2)^{4-d} 
|h|^2 
|q - q^{-1}|^2 
\int {\rm d}^d \r \, \non \\
&& \times \Bigg\{
\cG_0 (\r  | m_1^2)\cG_0 (\r  | m_2^2)\cG_0 (\r  | m_4^2) 
\Big( \cG_0 (\r  | m_2^2)
-\cG_0 (\r  | m_3^2) \Big)
\non\\
&&+ \frac{1}{4(N-1)} 
\cG_0 (\r  | 0) \Big( 2N \cG_0 (\r  | m_2^2)\cG_0 (\r  | m_3^2) 
+ (N-2) 
\big[\cG_0 (\r  | m_3^2)\big]^2 
\non \\
&&\qquad \qquad \qquad \qquad 
- (3N-2)
\big[\cG_0 (\r  | m_2^2)\big]^2 \Big)
 \Bigg\}~.
~~~
\eea
It remains to make use of the identity
\bea
(\m^2)^{4-d} 
\int {\rm d}^d \r \,
\cG_0 (\r  | m_1^2)\,\cG_0 (\r  | m_2^2)\,\cG_0 (\r  | m_3^2) 
=-I(m_1^2,m_2^2,m_3^2)~,
\label{propertime-->momentum}
\eea
where $I(x,y,z)$ is defined by (\ref{I(xyz)}).
As  a result, the K\"ahler potential takes the form
\bea
K(\f, \bar \f ) =  4 \frac{N-2}{N} |h|^2  |q - q^{-1}|^2 
\Big\{  I(m_1^2,m_2^2,m_4^2) &-& I(m_1^2,m_3^2,m_4^2)  
\label{KP} \\
+ \frac{1}{4(N-1)} \Big( 2N I(0,m_2^2,m_3^2) 
+  (N-2) I(0,m_3^2,m_3^2) 
&-&(3N-2) I(0,m_2^2,m_2^2)  \Big) \Big\} ~.~~~~~~
\non
\eea

The two-loop K\"ahler potential 
proves to be finite in the limit $d\to 4$, as it should be.
To see this, one can make use, e.g.,  
of eqs. (\ref{e-expan1}) and  (\ref{e-expan2}).
Then, by rewriting (\ref{KP}) in the form
\bea
K(\f, \bar \f ) =   \frac{N-2}{N(N-1)} |h|^2  |q - q^{-1}|^2 
\Big\{  4(N-1)\Big( I(m_1^2,m_2^2,m_4^2) &-& I(m_1^2,m_3^2,m_4^2) \Big) 
\non \\
+ (3N-2)\Big(I(0,m_2^2,m_3^2) &-& I(0,m_2^2,m_2^2) \Big) \\
-(N-2)\Big(I(0,m_2^2,m_3^2) &- &I(0,m_3^2,m_3^2) \Big)\Big\}~,~~
\non 
\eea
and setting $d=4$, we can represent $K(\f, \bar \f ) $
as follows:
\be
K(\f, \bar \f ) =  (m_3^2 -m_2^2) \,F\Big( \frac{m_2}{m_1},  \,\frac{m_3}{m_1}, \, \frac{m_4}{m_1}
\Big)~,
\ee
for some transcendental function $F$. From here and eq. (\ref{m1--4}), 
it follows $K(\f, \bar \f ) \propto \f \bar \f $.  It is also seen that $K(\f, \bar \f )$
disappears in the limit of vanishing deformation, $q\to \pm 1$.

\subsection{Evaluation of \mbox{$\bm{\Gamma_{\rm I + II}^{(A)} } $}}

Consider the contribution in the first line of (\ref{G:I+II-A1})
plus the one obtained by  $e \to -e$: 
\bea
\D \Gamma_1 &=&
 \frac{g^2}{2^9}ÊN(N-1)
  \int \rd^8 z \int \rd^8 z' \,
 G^{(0)} [\bar{\cD}^2, \cD^2] {\bf G}^{(e)} [\bar{\cD}'^2, \cD'^2] 
{\bf G}^{ \prime \, (e)} 
+ (e \leftrightarrow -e) \non \\
&=& 
\frac{g^2}{2^8}ÊN(N-1)
 \int \rd^8 z \int \rd^8 z' \, 
G^{(0)} [\bar{\cD}^2, \cD^2] {\bf G}^{(e)} [\bar{\cD}'^2, \cD'^2] 
{\bf G}^{ \prime \, (e)} ~.
\eea
Since $[\bar{\cD}'^2, \cD'^2] 
{\bf G}^{ \prime \, (e)}  =- [\bar{\cD}^2, \cD^2] {\bf G}^{(-e)}$, we can rewrite 
the above expression as 
\bea
\D \G_1 &=& -
\frac{g^2}{2^8}ÊN(N-1)
 \int \rd^8 z \int \rd^8 z' \, 
G^{(0)} (z,z') [\bar{\cD}^2, \cD^2] {\bf G}^{(e)}(z,z') 
[\bar{\cD}^2, \cD^2] {\bf G}^{(-e)}(z,z')~.
\non
\eea
Here the Green's function $G^{(0)} $ is massless, 
while the Green's functions ${\bf G}^{(e)}$ and  ${\bf G}^{(-e)}$
possess the same mass, $m_4$. Therefore, the evaluation
 of $\D \G_1$ can be carried out using the procedure employed in 
\cite{KM3,KM4}.

It follows from the explicit structure of the propagators, 
given in section 5,
that the integrand in $\D \G_1$ contains the contribution
\bea
&&\d^2(\z) \d^2(\bar \z )\,s^{-1} (\z-{\rm i} s\, e W) \r^a \s_a 
({\bar \z} +{\rm i}s\,e {\bar W}) \,t^{-1} 
(\z+{\rm i} t\, e W) \r^b \s_b 
({\bar \z} -{\rm i}t\, e {\bar W}) \non \\
&=&  e^4\, st\,\d^2(\z) \d^2(\bar \z )\, (W\r^a\s_a{\bar W})^2
=-\frac{e^4}{2} \,st\,\d^2(\z) \d^2(\bar \z )\, \r^2\,W^2 {\bar W}^2~.
\eea
Here the Grassmann delta-function, $\d^2(\z) \d^2(\bar \z )$, 
allows one to do the integral over ${\rm d}^4\q'$.
Changing bosonic integration variables, $x' \to \r$, 
one then obtains 
\bea
\D \G_1 &=& \hf e^4 N(N-1) g^2 \int {\rm d}^8z \,W^2 {\bar W}^2 \non \\
& \times &
\int\limits_0^\infty {\rm d}({\rm i}s)  {\rm d}({\rm i}t){\rm d}({\rm i}u)\, 
s t \int {\rm d}^4 \r \,\r^2\,
K_0 (\r , {\rm i}s | m_4^2) \, K_0 (\r , {\rm i}t | m_4^2)\, 
K_0 (\r , {\rm i}u | 0) ~.~~
\eea
The multiple integral in the second line can be readily evaluated.
The result is:
\bea
\D \G_1 &=& e^4 N(N-1) \frac{g^2}{3(4\p)^4}
\int {\rm d}^8z \,\frac{W^2 {\bar W}^2}{m_4^4} ~.
\eea

Consider the contribution in the second line of (\ref{G:I+II-A1})
plus the one  obtained by  $e \leftrightarrow -e$: 
\bea
\D \Gamma_2 &=&
 \frac{g^2}{2^9}ÊN(N-2)
  \int \rd^8 z \int \rd^8 z' \,G^{(e)}Ê\Big(
 [\bar{\cD}^2, \cD^2] {\bf G}^{(0)} [\bar{\cD}'^2, \cD'^2] 
{\bf G}^{ \prime \, (e)}
\non \\
&& \qquad \qquad
+[\bar{\cD}^2, \cD^2] {\bf G}^{(-e)} [\bar{\cD}'^2, \cD'^2] 
{\bf G}^{ \prime \, (0)} \Big) 
+ (e \leftrightarrow -e)~.
\eea
${}$For the background chosen, this reduces to 
\bea
\D \G_2 &=& -
\frac{g^2}{2^7}ÊN(N-2)
 \int \rd^8 z \int \rd^8 z' \, 
G^{(e)} (z,z') [\bar{\cD}^2, \cD^2] {\bf G}^{(0)}(z,z') 
[\bar{\cD}^2, \cD^2] {\bf G}^{(-e)}(z,z')~.
\non
\eea
The integrand can be seen to contain the contribution
\bea
&&\d^2 (\z-{\rm i} s\, e W) \d^2({\bar \z} +{\rm i}s\,e {\bar W})\,  
t^{-1} (\z+{\rm i} t\, e W) \r^a \s_a ({\bar \z} -{\rm i}t\, e {\bar W}) \,
u^{-1}  \z \r^b \s_b {\bar \z} \non \\
&=& - \frac{e^4}{2}\, \frac{s^2 (s+t)^2}{tu}
\d^2 (\z-{\rm i} s\, e W) \d^2({\bar \z} +{\rm i}s\,e {\bar W})\, 
\r^2\,W^2 {\bar W}^2~.
\eea
Here the shifted Grassmann delta-function, 
$\d^2 (\z-{\rm i} s\, e W) \d^2({\bar \z} +{\rm i}s\,e {\bar W})$,
can be used to do the integral over ${\rm d}^4\q'$.
Changing bosonic integration variables, $x' \to \r$, 
and then dimensionally continuing, ${\rm d}^4 \r \to {\rm d}^d \r$, 
we arrive at 
\bea
\D \G_2 &=& e^4 N(N-2) (\m^2)^{4-d} g^2
 \int {\rm d}^8z \,W^2 {\bar W}^2
\int\limits_0^\infty {\rm d}({\rm i}s)  {\rm d}({\rm i}t){\rm d}({\rm i}u)\, 
\frac{s^2 (s+t)^2}{tu} \non \\
&& \qquad \times 
\int {\rm d}^d \r \,\r^2\,
\cK_0 (\r , {\rm i}s | m_1^2) \,\cK_0 (\r , {\rm i}t | m_4^2)\, 
\cK_0 (\r , {\rm i}u | m_2^2) ~.
\eea
This can be rewritten as
\bea
\D \G_2 &=&- e^4 N(N-2) (\m^2)^{4-d} g^2
\int {\rm d}^8z \,W^2 {\bar W}^2
\int\limits_0^{{\rm i}\infty }{\rm d}\tilde{s} \, {\rm d}\tilde{t}\,
{\rm d}\tilde{u}\, 
\frac{\tilde{s}^2 (\tilde{s}+\tilde{t})^2}{\tilde{t}\tilde{u}} \non \\
&&\qquad \times 
\int {\rm d}^d \r \,\r^2\,
\cK_0 (\r , \tilde{s} | m_1^2) \,\cK_0 (\r , \tilde{t} | m_4^2)\, 
\cK_0 (\r , \tilde{u} | m_2^2) ~
\eea
where we have introduced  the notation
\bea
\tilde{s} = {\rm i}s~, \quad \tilde{t} = {\rm i}t~, \quad
\tilde{u} = {\rm i}u~.
\eea
To express this quantum correction in terms of two-loop momentum 
integrals, we may use integral  identities such as 
$$ 
0 =\int {\rm d}^d \r \, \frac{\pa}{\pa \r^a} \Big(\r^a\, f(\r^2)\Big)~,
$$
to obtain 
\bea
&&\int {\rm d}^d \r \,\frac{\r^2}{
\tilde{u}} \,
\cK_0 (\r , \tilde{s} | m_1^2) \,\cK_0 (\r , \tilde{t} | m_4^2)\, 
\cK_0 (\r , \tilde{u} | m_2^2) \non \\
&=&4 \int {\rm d}^d \r \,
\Big( \frac{d}{2} - \frac{ \r^2}{4\tilde{s}}
- \frac{ \r^2}{4\tilde{t}}\Big)\,
\cK_0 (\r , \tilde{s} | m_1^2) \,\cK_0 (\r , \tilde{t} | m_4^2)\, 
\cK_0 (\r , \tilde{u} | m_2^2) ~.
\eea
Using the explicit form of the heat kernel, eq. (\ref{d-kernel}), 
we can now represent 
\bea
&&\int {\rm d}^d \r \,\frac{\tilde{s}\r^2}{
\tilde{t} \tilde{u}} \,
\cK_0 (\r , \tilde{s} | m_1^2) \,\cK_0 (\r , \tilde{t} | m_4^2)\, 
\cK_0 (\r , \tilde{u} | m_2^2) \non \\
&=&-4 \int {\rm d}^d \r \,
\cK_0 (\r , \tilde{s} | m_1^2) \,
\cK_0 (\r , \tilde{u} | m_2^2) \,
\Big\{ \frac{d}{2} + (\tilde{s}+\tilde{t})
\big(m_4^2 + \frac{\pa}{\pa \tilde{t} } \big)\Big\}
\cK_0 (\r , \tilde{t} | m_4^2)~.
\eea
This result allows us to obtain
\bea
\D \G_2 &=&4 e^4 N(N-2) (\m^2)^{4-d} g^2
\int {\rm d}^8z \,W^2 {\bar W}^2 
\Bigg\{
\int\limits_0^{{\rm i}\infty }{\rm d}\tilde{s} \, {\rm d}\tilde{t}\,
{\rm d}\tilde{u}\, 
\tilde{s}(\tilde{s}+\tilde{t})^2
\non \\
& \times &
\Big( \frac{d}{2} -3 + m_4^2 (\tilde{s}+\tilde{t}) \Big)
\int {\rm d}^d \r \,
\cK_0 (\r , \tilde{s} | m_1^2) \,\cK_0 (\r , \tilde{t} | m_4^2)\, 
\cK_0 (\r , \tilde{u} | m_2^2)
\non \\
&&- \int\limits_0^{{\rm i}\infty }{\rm d}\tilde{s} \, 
\tilde{s}^4 \,\cK_0 (0 , \tilde{s} | m_1^2) \, 
\int\limits_0^{{\rm i}\infty }{\rm d}\tilde{u}\, \cK_0 (0 , \tilde{u} | m_2^2)
\Bigg\}~.
\eea

Similar calculations
can be applied to evaluate  the contribution 
in the third line of (\ref{G:I+II-A1}),
\bea
\D \Gamma_3 &=&
 \frac{g^2}{2^9}ÊN  \int \rd^8 z \int \rd^8 z' \,G^{(e)}Ê\Big(
 [\bar{\cD}^2, \cD^2] \tilde{{\bf G}}^{(0)} [\bar{\cD}'^2, \cD'^2] 
{\bf G}^{ \prime \, (e)}
\non \\
&& \qquad \qquad
+[\bar{\cD}^2, \cD^2] {\bf G}^{(-e)} [\bar{\cD}'^2, \cD'^2] 
\tilde{{\bf G}}^{ \prime \, (0)} \Big) 
+ (e \leftrightarrow -e)~.
\eea
One obtains
\bea
\D \G_3&=& 
4 e^4 N (\m^2)^{4-d} g^2
\int {\rm d}^8z \,W^2 {\bar W}^2 
\Bigg\{
\int\limits_0^{{\rm i}\infty }{\rm d}\tilde{s} \, {\rm d}\tilde{t}\,
{\rm d}\tilde{u}\, 
\tilde{s}(\tilde{s}+\tilde{t})^2
\non \\
& \times &
\Big( \frac{d}{2} -3 + m_4^2 (\tilde{s}+\tilde{t}) \Big)
\int {\rm d}^d \r \,
\cK_0 (\r , \tilde{s} | m_1^2) \,\cK_0 (\r , \tilde{t} | m_4^2)\, 
\cK_0 (\r , \tilde{u} | m_3^2)
\non \\
&&- \int\limits_0^{{\rm i}\infty }{\rm d}\tilde{s} \, 
\tilde{s}^4 \,\cK_0 (0 , \tilde{s} | m_1^2) \, 
\int\limits_0^{{\rm i}\infty }{\rm d}\tilde{u}\, \cK_0 (0 , \tilde{u} | m_3^2)
\Bigg\}~.
\eea

Making use of  the relations  (\ref{propertime-->momentum}) and 
 \be
\int\limits_0^{{\rm i}\infty }{\rm d}\tilde{s}\, 
 \cK_0 (0 , \tilde{s} | m^2) ={\rm i} \,J(m^2)~, 
\ee 
the results for $\D \G_2$ and $\D \G_3$  obtained 
can be transformed  to the final form:
\bea
\D \G_2 &=&-4 e^4 N(N-2) g^2
\int {\rm d}^8z \,W^2 {\bar W}^2 \non \\
&& \times 
\Bigg\{ \frac{\pa}{\pa m_1^2} \Big( \frac{\pa}{\pa m_1^2} +\frac{\pa}{\pa m_4^2} \Big)^2
\Big[ 3-\frac{d}{2} +m_4^2 \Big( \frac{\pa}{\pa m_1^2} +\frac{\pa}{\pa m_4^2} \Big) \Big]
I(m_1^2,m_2^2,m_4^2)~~~~ \non \\
&&\qquad \qquad - \Big( \frac{\pa}{\pa m_1^2}\Big)^4 J(m_1^2) J(m_2^2) \Bigg\}~,
\label{dG2-final}\\
\D \G_3 &=&-4 e^4 N 
g^2
\int {\rm d}^8z \,W^2 {\bar W}^2 \non \\
&& \times 
\Bigg\{ 
\frac{\pa}{\pa m_1^2} \Big( \frac{\pa}{\pa m_1^2} +\frac{\pa}{\pa m_4^2} \Big)^2
\Big[ 3-\frac{d}{2} +m_4^2 \Big( \frac{\pa}{\pa m_1^2} +\frac{\pa}{\pa m_4^2} \Big) \Big]
I(m_1^2,m_3^2,m_4^2)~~~~ \non \\
 &&\qquad \qquad - \Big( \frac{\pa}{\pa m_1^2}\Big)^4 J(m_1^2) J(m_3^2) \Bigg\}~.
\label{dG3-final}
\eea

${}$Finally, the expression in the fourth line of (\ref{G:I+II-A1})
does not produce any contribution to the effective action, 
since all the Green's functions appearing in it are neutral, and therefore
do not couple to the background vector multiplet.

\subsection{Evaluation of \mbox{$\bm{ \Gamma_{\rm I + II}^{(B)} }$}}
Consider the expression 
in the first line of (\ref{G:I+II-B1})
plus the one obtained by  $e \to -e$: 
\bea
&&
-\frac{1}{2^8} \,Ê 
\frac{N-2}{N} 
 |h|^2  |q - q^{-1}|^2 
\int \rd^8 z 
\rd^8 z'Ê \, 
G^{(e)} 
\big( \hat{{\bf G}}^{(0)} 
\check{{\bf G}}^{\prime \, (e)}ÊÊ + 
\hat{{\bf G}}^{(-e)}
 \check{{\bf G}}^{\prime \, (0)} \big) 
+(e \leftrightarrow -e) \nonumber \\
&=&
-\frac{1}{2^6} \,Ê 
\frac{N-2}{N} |h|^2  
|q - q^{-1}|^2 
\int \rd^8 z 
\rd^8 z'Ê \, 
G^{(e)} (z,z')\,
\bar{\cD}^2 \cD'^2 {\bf G}^{(0)} (z,z')\,
\bar{\cD}^2 \cD'^2 {\bf G}^{(e)} (z,z')
\non  \\
&=&
-\frac{1}{2^6} \,Ê 
\frac{N-2}{N} |h|^2  
|q - q^{-1}|^2 
\int \rd^8 z 
\rd^8 z'Ê \, 
 {\bf G}^{(0)} (z,z')\,
\bar{\cD}^2 \cD^2G^{(e)} (z,z')\,
\bar{\cD}^2 \cD^2 {\bf G}^{(e)} (z,z')~.
\non
\eea
Ignoring the $W$-independent quantum correction, 
which contributes to the K\"ahler potential, eqs.
(\ref{Gcl1}) and (\ref{Gcl2}) give
\bea
\D \G_4 &=&-4 \,Ê 
\frac{N-2}{N} (\m^2)^{4-d}
 |h|^2  |q - q^{-1}|^2 
\int \rd^8 z  \int {\rm d}^d \r 
\non \\
&&\times 
\int\limits_0^{{\rm i}\infty }{\rm d}\tilde{s} \, {\rm d}\tilde{t}\,
{\rm d}\tilde{u}\, 
\Big[ \exp \Big\{ \frac{1}{6} e^4 (\tilde{s}^3 +\tilde{t}^3)
W^2 {\bar W}^2 -\frac{\rm i}{2} e^2
(\tilde{s} +\tilde{t}) \r^a W\s_a{\bar W} \Big\}
-1\Big]
\non \\
&& \times 
\cK_0 (\r , \tilde{s} | m_1^2) \,\cK_0 (\r , \tilde{t} | m_4^2)\, 
\cK_0 (\r , \tilde{u} | m_2^2)~.
\eea
Since $W^3=0$, this is equivalent to 
\bea
\D \G_4 &=&-2 e^4\,Ê 
\frac{N-2}{N} (\m^2)^{4-d}
 |h|^2  |q - q^{-1}|^2 
\int \rd^8 z  \,W^2 {\bar W}^2 
\int {\rm d}^d \r 
\int\limits_0^{{\rm i}\infty }{\rm d}\tilde{s} \, {\rm d}\tilde{t}\,
{\rm d}\tilde{u}\, 
\non \\
&& \times
 \Big\{ \frac{1}{3} (\tilde{s}^3 +\tilde{t}^3)
+ \frac{1}{8}(\tilde{s} +\tilde{t} )^2 \r^2 \Big\}
\cK_0 (\r , \tilde{s} | m_1^2) \,\cK_0 (\r , \tilde{t} | m_4^2)\, 
\cK_0 (\r , \tilde{u} | m_2^2)~.~~
\label{D-G4-1}
\eea
Representing 
\bea
\frac{1}{4} \r^2 \,\cK_0 (\r , \tilde{u} | m_2^2)
&=&\frac{\pa}{\pa \tilde{u} } \Big\{ \tilde{u}^2 \cK_0 (\r , \tilde{u} | m_2^2) \Big\} \non \\
&& + m_2^2 \tilde{u}^2 \cK_0 (\r , \tilde{u} | m_2^2)
+\Big(\frac{d}{2} -2 \Big) \tilde{u}\,\cK_0 (\r , \tilde{u} | m_2^2)~,
\eea
we obtain 
\bea
\D \G_4 &=&-2 e^4\,Ê 
\frac{N-2}{N} (\m^2)^{4-d}
 |h|^2  |q - q^{-1}|^2 
\int \rd^8 z  \,W^2 {\bar W}^2 
\int {\rm d}^d \r 
\int\limits_0^{{\rm i}\infty }{\rm d}\tilde{s} \, {\rm d}\tilde{t}\,
{\rm d}\tilde{u}\, 
\non \\
&& \times
 \Big\{ \frac{1}{3} (\tilde{s}^3 +\tilde{t}^3)
+ \hf m_2^2(\tilde{s} +\tilde{t} )^2 \tilde{u}^2 \Big\}
\cK_0 (\r , \tilde{s} | m_1^2) \,\cK_0 (\r , \tilde{t} | m_4^2)\, 
\cK_0 (\r , \tilde{u} | m_2^2)~.~~
\label{D-G4-2}
\eea
Here we have used the fact that 
\bea
\lim_{d\to 4} \,
(d -4)
\int {\rm d}^d \r 
\int\limits_0^{{\rm i}\infty }{\rm d}\tilde{s} \, {\rm d}\tilde{t}\,
{\rm d}\tilde{u}\, 
(\tilde{s} +\tilde{t} )^2 \tilde{u}\,
\cK_0 (\r , \tilde{s} | m_1^2) \,\cK_0 (\r , \tilde{t} | m_4^2)\, 
\cK_0 (\r , \tilde{u} | m_2^2) =0~,
\non
\eea
since the integral can be seen to be  finite.

The $F^4$ quantum correction generated by 
the expression in the second  line of (\ref{G:I+II-B1})
is obtained from (\ref{D-G4-2}) by 
changing the overall sign and 
replacing $m_2$ by $m_3$:
\bea
\D \G_5 &=&2 e^4\,Ê 
\frac{N-2}{N} (\m^2)^{4-d}
 |h|^2  |q - q^{-1}|^2 
\int \rd^8 z  \,W^2 {\bar W}^2 
\int {\rm d}^d \r 
\int\limits_0^{{\rm i}\infty }{\rm d}\tilde{s} \, {\rm d}\tilde{t}\,
{\rm d}\tilde{u}\, 
\non \\
&& \times
 \Big\{ \frac{1}{3} (\tilde{s}^3 +\tilde{t}^3)
+ \hf m_3^2(\tilde{s} +\tilde{t} )^2 \tilde{u}^2 \Big\}
\cK_0 (\r , \tilde{s} | m_1^2) \,\cK_0 (\r , \tilde{t} | m_4^2)\, 
\cK_0 (\r , \tilde{u} | m_3^2)~.~~
\label{D-G5-2}
\eea

The results for $\D \G_4$ and $\D \G_5$  obtained 
can be transformed  to the final form:
\bea
\D \G_4 &=&-\frac{2}{3} e^4\,Ê 
\frac{N-2}{N} 
 |h|^2  |q - q^{-1}|^2 
\int \rd^8 z  \,W^2 {\bar W}^2 \non \\
&& \times 
\Bigg\{
 \Big( \frac{\pa}{\pa m_1^2}\Big)^3
+ \Big( \frac{\pa}{\pa m_4^2}\Big)^3
 - \frac{3}{2} m_2^2\Big( \frac{\pa}{\pa m_2^2}\Big)^2
\Big( \frac{\pa}{\pa m_1^2} +\frac{\pa}{\pa m_4^2} \Big)^2
\Bigg\}
I(m_1^2,m_2^2,m_4^2)~,~~~ ~~~
\label{dG4-final} \\
\D \G_5 &=&-\frac{2}{3} e^4\,Ê 
\frac{N-2}{N} 
 |h|^2  |q - q^{-1}|^2 
\int \rd^8 z  \,W^2 {\bar W}^2 \non \\
&& \times 
\Bigg\{
 \Big( \frac{\pa}{\pa m_1^2}\Big)^3
+ \Big( \frac{\pa}{\pa m_4^2}\Big)^3
 - \frac{3}{2} m_2^2\Big( \frac{\pa}{\pa m_3^2}\Big)^2
\Big( \frac{\pa}{\pa m_1^2} +\frac{\pa}{\pa m_4^2} \Big)^2
\Bigg\}
I(m_1^2,m_3^2,m_4^2)~.~~~ ~~~
\label{dG5-final} 
\eea

The expressions 
in the third  and  fourth lines of (\ref{G:I+II-B1})
do not produce any quantum corrections, since they involve
neutral  Green's functions decoupled from the background vector multiplet.

\subsection{Evaluation of \mbox{$\bm{ \Gamma_{\rm III} }$}}
Let us turn to the evaluation of $\G_{\rm III}$, eq. (\ref{G-III}).
As is obvious from the structure of the superpropagators, 
the expression in the last line of  (\ref{G-III}) does not contribute.
The expression in the first line of (\ref{G-III}), plus the one obtained by 
$e \to -e$, 
can be seen to generate a finite quantum correction.
It has the form:
\bea
\D \G_6 &=& -2e^4
N(N-1)\int \rd^8 z  \, m_4^2
W^2 {\bar W}^2 \non \\
&&\times 
\int\limits_0^{{\rm i}\infty }{\rm d}\tilde{s} \, {\rm d}\tilde{t}\,
{\rm d}\tilde{u}\, 
\tilde{s}^2\,\tilde{t}^2
\int {\rm d}^4 \r \,
K_0 (\r , \tilde{s} | m_4^2) \, K_0 (\r , \tilde{t} | m_4^2)\, 
K_0 (\r , \tilde{u} | 0)~.~~
\eea
Here one of the kernels is massless, and the others possess 
the same mass.
Therefore, the evaluation
 of $\D \G_6$ can be carried out using the procedure employed in 
\cite{KM3,KM4}. The result is
\bea
\D \G_6 &=& e^4 N(N-1)
\frac{2}{3(4\p)^4}
\int \rd^8 z  \, \frac{W^2 {\bar W}^2}{m_4^4}~.
\eea

The expressions in the second and third  lines of (\ref{G-III})
lead to 
\bea
\D \G_7 &=& 2e^4\frac{(N-2)^2}{N-1} (\m^2)^{4-d} 
 |h|^2 |q - q^{-1}|^2
\int \rd^8 z  \,\f{\bar \f}\,W^2 {\bar W}^2
\int\limits_0^{{\rm i}\infty }{\rm d}\tilde{s} \, {\rm d}\tilde{t}\,
{\rm d}\tilde{u}\, 
\tilde{s}^2\, (\tilde{s} + \tilde{t})^2
\non \\
&&\times
\int {\rm d}^d \r \,
\cK_0 (\r , \tilde{s} | m_1^2) \,\cK_0 (\r , \tilde{t} | m_4^2)\, 
\Big( \cK_0 (\r , \tilde{u} | m_2^2) -\cK_0 (\r , \tilde{u} | m_3^2)\Big)~.
\eea

The result for $\D \G_7$   obtained 
can be transformed  to the final form:
\bea
\D \G_7 &=& -2e^4\frac{(N-2)^2}{N-1} 
 |h|^2 |q - q^{-1}|^2
\int \rd^8 z  \,\f{\bar \f}\,W^2 {\bar W}^2
 \non \\
&&\times 
 \Big( \frac{\pa}{\pa m_1^2}\Big)^2
\Big( \frac{\pa}{\pa m_1^2} +\frac{\pa}{\pa m_4^2} \Big)^2
\Bigg\{
I(m_1^2,m_2^2,m_4^2)- I(m_1^2,m_3^2,m_4^2)\Bigg\}~.
\label{dG7-final} 
\eea

\subsection{Evaluation of \mbox{$\bm{ \Gamma_{\rm IV} }$}}
It follows from (\ref{G-IV})
\bea
\Gamma_{\rm IV} &=&  \frac{1}{8}N g^2  \int \rd^8 z \lim_{z' \rightarrow z} 
 G^{(e)} \Big( \hat{\tilde{{\bf G}}}^{(0)}  
+ (N-2) \hat{{\bf G}}^{(0)} + (N-1) \hat{{\bf G}}^{(e)}  \Big)
 \nonumber \\
&=& 2e^4 N(\m^2)^{4-d} \int \rd^8 z  \,W^2 {\bar W}^2
\int\limits_0^{{\rm i}\infty }{\rm d}\tilde{s} \, \tilde{s}^4\,
\cK_0 (0 , \tilde{s} | m_1^2) \non \\
&&\times  \int\limits_0^{{\rm i}\infty }{\rm d}\tilde{u} 
\Big\{ \cK_0 (0 , \tilde{u} | m_3^2) +(N-2)\cK_0 (0 , \tilde{u} | m_2^2) 
+(N-1) \cK_0 (0 , \tilde{u} | m_4^2) \Big\}~.
\eea
This result can equivalently be rewritten as follows:
\bea
\Gamma_{\rm IV} &=&  -2e^4 N
 \int \rd^8 z  \,W^2 {\bar W}^2 \non \\
&&  \times  \Big( \frac{\pa}{\pa m_1^2}\Big)^4 J(m_1^2) 
\Bigg\{ J(m_3^2) +(N-2) J(m_2^2) +(N-1)J(m_4^2)\Bigg\}~.
\label{GIV-final} 
\eea

\subsection{Cancellation of divergences}
We conclude this paper by demonstrating that the two-loop effective action is finite. 
More precisely, we demonstrate the cancellation of all divergent $F^4$ contributions.
This only requires the use of eqs. (\ref{e-expan1}) and  (\ref{e-expan2}), 
along with the well-known expression for the divergent part of 
the one-loop tadpole $J(x)$:
\be
(4\p)^2 \, J_{\rm div}(x) = -\frac{1}{\e}\, x~.
\ee

Let us first consider the figure-eight contribution, eq. (\ref{GIV-final}).
Its divergent part is
\bea
(\Gamma_{\rm IV})_{\rm div} &=&\frac{1}{\e} \, \frac{4e^4 N}{(4\p)^4 }
 \int \rd^8 z  \,
\frac{W^2 {\bar W}^2}{m_1^3}
\left\{m_3^2 +(N-2) m_2^2 +(N-1)m_4^2 \right\}~.
\eea
Making use of relations  (\ref{m1--4}) gives 
\be
m_3^2 +(N-2) m_2^2 +(N-1)m_4^2 =(N-1) m_1^2~,
\label{massid}
\ee
and therefore
\bea
(\Gamma_{\rm IV})_{\rm div} &=&\frac{1}{\e} \, \frac{4e^4 N(N-1)}{(4\p)^4 }
 \int \rd^8 z  \,
\frac{W^2 {\bar W}^2}{m_1^2}~.
\eea
This coincides with the expression for $(\Gamma_{\rm IV})_{\rm div} $ 
that occurs in the undeformed $\cN=4$ SYM theory \cite{KM3,KM4}.

Now, let us turn to the quantum corrections $\D \G_{1}, \dots, \D \G_7$ 
produced by the sunset supergraphs. Here  $\D \G_{1}$ and $\D \G_{6}$ are 
finite. Using eqs.  (\ref{e-expan1}) and  (\ref{e-expan2}), one can explicitly 
check that  the contribution $\D \G_{4}+\D \G_{5}$, which is defined by  eqs.
(\ref{dG4-final}) and (\ref{dG5-final}), is finite, 
and so is $\D \G_7$, eq. (\ref{dG7-final}). Therefore, it remains to analyze 
the quantum corrections $\D \G_{2}$ and $\D \G_{3}$ given by eqs.
(\ref{dG2-final}) and (\ref{dG3-final}). Their direct inspection, with the use of 
(\ref{massid}), gives
\be
\Big(\D \G_{2} +\D \G_{3}\Big)_{\rm div} = -(\Gamma_{\rm IV})_{\rm div} ~.
\ee
Therefore, the two-loop effective action is free of ultraviolet divergences. 
${}$
\\

\noindent
{\bf Acknowledgements:}\\
One of us (S.M.K.) is grateful to Gerald Dunne for  pointing out 
important references, and to Arkady Tseytlin for helpful discussions and hospitality
at Imperial College. 
This work  was supported  
by the Australian Research Council and by a UWA research grant.

\begin{appendix}

\sect{Integral representation for \mbox{$\bm{I(x,y,z)}$}}
\label{two-loop-int}
In \cite{FJJ}, a useful integral representation for
the completely symmetric function
\be
I(x,y,z) = \frac{(\m^2)^{4-d} }{(2\p)^{2d} }
\int \frac{ {\rm d}^d k \, {\rm d}^d q}{ (k^2 +x) (q^2 +y)
((k+q)^2 +z) }~, 
\qquad d=4-2\e
\label{I(xyz)}
\ee
was obtained using 
the differential equations method \cite{Kotikov} and 
the method of characteristics (see, e.g., \cite{CH}).
In that work, only the case $\D(x,y,z)<0$ was treated in detail.
AsÊ noted earlier, two-loop contributions to the effective action
in $\b$-deformed theories correspond to the case $\D(x,y,z)>0.$
For completeness, we provide a detailed derivation of a representation
for $I(x,y,z)$ in this case.

Using the integration-by-parts  
technique \cite{CT},
the identity
\be
0 = \intÊ {\rm d}^d k \, {\rm d}^d q \, \frac{\partial}{\partial k_{\m} } \Big\{
\frac{k_{\m}}{ (k^2 +x) (q^2 +y)
((k+q)^2 +z) } \Big\}
\label{total-deriv}
\ee
can be seen 
to be  equivalent
to the following 
differential equation:
\bea 
0&=& \Big( d-3 -2x\, \frac{\pa}{\pa x} +(y-x-z)\, \frac{\pa}{\pa z} \Big) 
I(x,y,z) - J'(z) \big(J(x) -J(y) \big)~, 
\label{pde1}
\eea
with $J(x)$ the tadpole integral (\ref{tadpole}) satisfying
the first-order equation
\be
J'(x)  = \frac{d-2}{2} \, \frac{J(x)}{x}~.
\ee
Making use of two more equations that follow 
from (\ref{pde1}) by applying cyclic permutations 
of $x$, $y$ and $z$, 
one can 
establish the following differential equation \cite{FJJ} for $I(x,y,z):$
\bea
0 &=& \Big[ (y-z) \frac{\partial}{\partial x} 
+ (z-x)Ê \frac{\partial}{\partial y} + (x-y)Ê \frac{\partial}{\partial z} \Big] I(x,y,z) \nonumber \\
&+& J'(z) \big( J(x) - J(y) \big) + J'(x) \big( J(y) - J(z) \big) + J'(y) \big( J(z) - J(x) \big)~.
\eea
In \cite{FJJ}, it was recognized that this equation
can be solved by the method of characteristics.
By introducing a one-parameter flow $(x(t), y(t), z(t))$
in the parameter space of masses such that
\be
\frac{{\rm d} x(t)}{{\rm d}t} = y(t) - z(t)~, \quad
\frac{{\rm d} y(t)}{{\rm d} t} = z(t) - x(t)~,
\quad \frac{{\rm d} z(t)}{{\rm d}t} = x(t) - y(t)~,
\label{flow}
\ee
and using the expression (\ref{tadpole}) for the one-loop tadpole,
equation (\ref{I(xyz)}) becomes
\be
\frac{{\rm d}}{{\rm d}t} I(x(t), y(t), z(t)) =Ê \G' \,x(t)^{\frac{d}{2} -2} \big( y(t)^{\frac{d}{2} -1}
- z(t)^{\frac{d}{2}-1} \big) + \, \, {\rm cyclic}~,
\label{floweqn}
\ee
with
\be
\G' = \frac{(\m^2)^{4-d} }{(4 \p)^{d} } \,Ê \G(1- \frac{d}{2}) \, \G(2-\frac{d}{2} ) 
= -\frac{d-2}{2} \Big[ \frac{(\m^2)^{2-d/2} }{(4 \p)^{d/2} } \,\G(1- \frac{d}{2})
\Big]^2
 ~.
\label{Gamma'}
\ee
If the endpoints of the flow are $(x(0), y(0), z(0)) = (X,Y,Z)$ and
$(x(1), y(1), z(1)) = (x,y,z),$
then integrating (\ref{floweqn}) yields
\be
I(x,y,z)Ê -Ê I(X,Y,Z) =Ê  \G'Ê \int_0^1 \rd t \,Ê \left[ x(t)^{\frac{d}{2} -2}
\big( y(t)^{\frac{d}{2} -1} - z(t)^{\frac{d}{2}-1} \big) + \, \, {\rm cyclic} \right] ~ .
\label{flowint}
\ee
The flow (\ref{flow}) preserves the values of
\be c \equiv x(t) + y(t) + z(t) \,\, {\rm and} \, \, \Delta \equiv \Delta(x(t),y(t),z(t))~,
\label{flowconstants}
\ee
and so, for a given endpoint $(x,y,z),$Ê the starting point $(X,Y,Z)$Ê
cannot be chosen arbitrarily. Nevertheless, the key point of \cite{FJJ}
is that it is possible to choose $(X,Y,Z)$ in such a way that
the integration constant $I(X,Y,Z)$ is a simpler integral
which can be determined in closed form.

Multiplying out the integrand in (\ref{flowint}), the equation can be rearranged as
\bea
I(x,y,z)Ê -Ê I(X,Y,Z) &=& - \, \G'ÊÊ \int_0^1 \rd tÊ \,
\frac{{\rm d} x(t)}{{\rm d}t} \, \left[ \big( y(t) z(t) \big)^{\frac{d}{2} -2}Ê + \, \,
{\rm cyclic} \right] ~.
\eea
Using the definitions (\ref{flowconstants}) for the constants $c$ and $\Delta,$
it is easy to establish that
\be
y(t) z(t) = \big(x(t) - \frac{c}{2} \big)^2 + \frac{\Delta}{4}~.
\ee
Therefore,
\be
I(x,y,z)Ê -Ê I(X,Y,Z) = - \, \G'ÊÊ \Big( \int_{X-\frac{c}{2}}^{x-\frac{c}{2}}
+Ê \int_{Y-\frac{c}{2}}^{y-\frac{c}{2}} +Ê \int_{Z-\frac{c}{2}}^{z-\frac{c}{2}} \Big) \, \rd s \,
\big(s^2 + \frac{\Delta}{4} \big)^{\frac{d}{2} - 2}~,
\label{integratedflow}
\ee
allowing the two-loop integral $I(x,y,z)$ to be expressed in terms of
an integral $I(X,Y,Z)$ which may be more easily evaluated.

In the caseÊ $\Delta(x,y,z) < 0$ considered in detail in \cite{FJJ},
the flow can be chosen to start at the point $(X,Y,0)$
-- the constants (\ref{flowconstants}) of the flow are
$c = X+Y = x+y+z$ and $\Delta = - (X-Y)^2 = \Delta(x,y,z),$ which can be solved
for $X$ and $Y$. Thus the integrated flow equation (\ref{integratedflow})
allowsÊ $I(x,y,z),$ with three nonvanishing masses, to be expressed in terms of
an integral $I(X,Y,0)$ with one vanishing mass. This integral in turn satisfies
the differential equation
\bea 0 &=& (X-Y) \frac{\partial I(X,Y,0)}{\partial X} +Ê (X-Y) \frac{\partial I(X,Y,0)}{\partial Y } \nonumber \\
&- & J'(X) \big( J(Y) - J(0) \big) + J'(Y) \big( J(X) - J(0) \big)~.
\eea
Again, this equation can be integrated using the method of
characteristicsÊ to yield \cite{FJJ}
\be
I(X,Y,0) - I(X-Y,0,0) = \Gamma' \, \int^{\frac{X+Y}{2}}_{\frac{X-Y}{2}} \rd s \, \big(s^2 - \frac{(X-Y)^2}{4} \big)^{\frac{d}{2} - 2}~.
\ee
This time, the integration constant is known in closed form,
and so the process is complete. The final result \cite{FJJ} is
\be
I(x,y,z) = I(\sqrt{- \Delta},0,0) + \Gamma' \, \big( F(\frac{c}{2} - y)
+ F(\frac{c}{2} - z) - F(x - \frac{c}{2} ) \big)~,
\ee
where
\be F(w) = \int_{\frac{\sqrt{- \D}}{2}}^{w} \rd s \, \big(s^2 + \frac{\Delta}{4} \big)^{\frac{d}{2} - 2}
\ee
and
\be
I(x,0,0) = \frac{(\mu^2)^{4-d}}{(4 \pi)^{d}} \,
\frac{\G(2 - \frac{d}{2}) \G(3-d) \G(\frac{d}{2} - 1)^2}{\G(\frac{d}{2})} \, x^{d-3}~.
\ee

An expression for $I(x,y,z)$ is given without derivation in \cite{FJJ} for the case
$\Delta(x,y,z) > 0.$  For completeness, we fill this gap here,
as it is the case of interest for $\b$-deformed theories. It is clear that the starting point
for the flow (\ref{flow}) in the parameter space of masses cannot be chosen to be $(X,Y,0),$Ê
as then $\Delta = - (X-Y)^2 $ is manifestly negative. Instead, for positive $\Delta,$
it is convenient to choose the starting point of the flow to correspond to two equal masses,
\be (x(0), y(0), z(0)) = (X,Y,Y)~.
\ee
The parameters $X$ and $Y$ are related to the constants (\ref{flowconstants}) by
\be
\D = 4 X Y - X^2, \quad c = X + 2 Y~.
\label{consts}
\ee
It can be checked that these equations admit solutions for $X$ and $Y$ which are real
and non-negative, as required for physical consistency, sinceÊ $X$ and $Y$ are the squares
of masses.\footnote{It is of interest to note that physical solutions do not exist
for $\Delta(x,y,z) < 0;$ it is not possible to ensure that $Y$ is non-negative.}
The integrated flow equation (\ref{integratedflow}) yields
\bea
I(x,y,z) &=& I(X,Y,Y) + \Gamma' \, \big(G(\frac{c}{2} - x ) + G(\frac{c}{2} - y ) + G(\frac{c}{2} - z ) \nonumber \\
&+& G(\frac{X}{2} - Y) - 2 \, G(\frac{X}{2}) \big)~,
\label{I-arbitrary}
\eea
with
\be
G(w) = \int_0^{w} \rd s \, \big(s^2 + \frac{\Delta}{4} \big)^{\frac{d}{2} - 2}~.
\label{G}
\ee

The integralÊ $I(X,Y,Y)$ can also be determined using the method of characteristics.
It satisfies the differential equation
\be
0 = X \, \frac{\partial I(X,Y,Y)}{\partial X} + \big( \frac{X}{2} - Y \big) \, \frac{\partial I(X,Y,Y)}{\partial Y}
+ J'(X) \big(J(X) - J(Y) \big)~.
\ee
Introducing a flow $(X(t), Y(t))$ satisfying
\be
\frac{{\rm d} X(t)}{{\rm d}t } = X(t)~, \qquad
\frac{{\rm d} Y(t)}{{\rm d}t} = \frac{X(t)}{2} - Y(t)~,
\label{flow2}
\ee
and using the expressions (\ref{tadpole}) for the one-loop tadpole, it follows that
\be
\frac{{\rm d}}{{\rm d}t} I(X(t), Y(t), Y(t))
= \G' \, Y(t)^{(\frac{d}{2} - 2)} \, \big( X(t)^{(\frac{d}{2} - 1)}
- Y(t)^{(\frac{d}{2} - 1)}Ê \big)~.
\label{flowdiff2}
\ee
This equation can be integrated from some convenient starting point $(X(0), Y(0))$
to the point $(X(1), Y(1)) = (X,Y)$ in order to yield an expression for $I(X,Y,Y)$ in terms of
a potentially simpler two-loop integral $I(X(0), Y(0), Y(0)).$
The flow (\ref{flow2}) preserves the value of $4 X(t) Y(t) - X(t)^2,$
which at the point $t=1$ takes the value
$4 X Y - X^2 = \D,$ as determined in (\ref{consts}).
The equation $\D = 4 X(t) Y(t) - X(t)^2,$ combined with (\ref{flow2}), allows us to rewrite
\bea
Y(t)^{(\frac{d}{2} - 2)} \, X(t)^{(\frac{d}{2} - 1)} &=&4^{(\frac{d}{2} - 2)} \,
\frac{{\rm d} X(t)}{{\rm d}t} \big( X(t) + \D \big)^{(\frac{d}{2} - 2)} , \nonumber \\
Y(t)^{(\frac{d}{2} - 2)} \, Y(t)^{(\frac{d}{2} - 1)}
&=& \big( \frac12 \frac{{\rm d} X(t)}{{\rm d}t} - \frac{{\rm d} Y(t)}{{\rm d}t} \big) \,
\left( \big( \frac{X(t)}{2} - Y(t) \big)^2 + \frac{\D}{4} \right)^{(\frac{d}{2} - 2)} ~.
~~~~
\eea
Substituting this into (\ref{flowdiff2}) and integrating from $t=0$ to $t=1,$
\bea
I(X,Y,Y) - I(X(0), Y(0), Y(0) ) &=& \G' \, \left( 2 \int_{\frac{X(0)}{2}}^{\frac{X}{2}} - \int_{\frac{X(0)}{2}- Y(0)}^{\frac{X}{2} - Y} \right) \, \rd s \, \big( s^2 + \frac{\D}{4} \big)^{(\frac{d}{2} - 2)}Ê \nonumber \\
&=& \G' \, \left( 2G(\frac{X}{2}) - G(\frac{X}{2} - Y ) -Ê 2G(\frac{X(0)}{2}) \right. \nonumber \\ &+& \left. G(\frac{X(0)}{2} - Y(0))Ê \right)~.
\label{flowint2}
\eea

Although it would be convenient to choose $(X(0), Y(0)) = (\tilde{X},0),$ so that
the integration constant in (\ref{flowint2}) would be $I(\tilde{X},0,0),$ the pointÊ $(\tilde{X},0)$
does not lie on the flow with $(X(1),Y(1)) = (X,Y),$ which is characterized by
$4 X(t) Y(t) - X(t)^2 = \D >0.$ The issue is that for the point $(\tilde{X},0),$
$4 X(t) Y(t) - X(t)^2 = - \tilde{X}^2,$ which is manifestly negative,
and therefore corresponds to a different flow. A suitable starting point for
the flow isÊÊ $(X(0), Y(0)) = (\tilde{X},\tilde{X}),$ for which $4 X(t) Y(t) - X(t)^2 = 3 \tilde{X}^2,$
which is manifestly positive and can therefore be equated to $\D,$ yielding
$\tilde{X} = \sqrt{\frac{\D}{3}}.$

Combining (\ref{I-arbitrary}) and (\ref{flowint2}), we find that for $\D>0,$
\be
I(x,y,z) = I( \sqrt{\frac{\D}{3}},Ê \sqrt{\frac{\D}{3}}, \sqrt{\frac{\D}{3}}) + \G' \, \big( G(\frac{c}{2} - x) +Ê G(\frac{c}{2} - y)
+ G(\frac{c}{2} - z) - 3 G(\frac{1}{2} \sqrt{\frac{\D}{3}}) \big)~.
\label{Inotfinal}
\ee
The integration constant $ I( \sqrt{\frac{\D}{3}}, \sqrt{\frac{\D}{3}}, \sqrt{\frac{\D}{3}})$ Ê 
is a two-loop integral with three equal masses, for which a variety of equivalent expressions 
exist in the literature. We choose the representation
\bea
I(\sqrt{\frac{\D}{3}}, \sqrt{\frac{\D}{3}}, \sqrt{\frac{\D}{3}}) &=& \frac{(\m^2)^{4-d} }{(2\p)^{2d} } 
\, \left(\frac{\D}{3} \right)^{\frac{d}{2}- \frac32} \, \left[Ê 3^{\frac12-d} 
\frac{2 \pi \, \G(5-d)}{(4-d) (d-2)(d-3)} \right. \nonumber \\
&+& \left.Ê \frac{6 \, \G(3 - \frac{d}{2})^2}{ (d-4)^2 (\frac{d}{2} -1)} \, {}_2F_1( 2 - \frac{d}{2}, 1 ; \frac32; \frac14) \right]
\label{3x}
\eea
given inÊ \cite{Broadhurst:1993mw} (based on results in \cite{FJ,FJJ,DST}).
The term involving the hypergeometric in (\ref{3x}) precisely cancels the contribution 
$ - 3 G(\frac{1}{2} \sqrt{\frac{\D}{3}}) $ in (\ref{Inotfinal}). To see this, by making 
the change of variable $s = \frac{\D}{4} \, u$ in (\ref{G}),
\bea
G(\frac{1}{2} \sqrt{\frac{\D}{3}}) &=& \frac{1}{2 \sqrt{3}} \, 
\left( \frac{\D}{4} \right)^{\frac{d}{2}- \frac32} \int_0^1 \rd u \, u^{- \frac12} \, ( 1 - \frac13 u)^{\frac{d}{2}-2} 
\nonumber \\
&=& \frac{1}{\sqrt{3}} \, \left( \frac{\D}{4} \right)^{\frac{d}{2}- \frac32} \, 
{}_2F_1(2 - \frac{d}{2}, \frac12; \frac32 ; - \frac13)
\eea
using (see, e.g., \cite{PBM})
\be
\int_0^1 \rd t \, t^{b-1} (1-t)^{c-b-1} (1-zt)^{-a} = \frac{\G(b) \G(c-b)}{\G(c)} \, {}_2F_1(a,b;c;z)~.
\ee
Applying the identity  \cite{PBM}
\be
{}_2F_1(a,b;c;z) = (1 - z)^{-a} \, {}_2F_1(a , c - b ; c; \frac{z}{z-1})
\ee
yields
\be
G(\frac{1}{2} \sqrt{\frac{\D}{3}})Ê = \frac12 \,\left( \frac{\D}{3} \right)^{\frac{d}{2}- \frac32}\, {}_2F_1(2 - \frac{d}{2} , 1 ; \frac32 ; \frac14),
\ee
from which the result follows.

As a result, (\ref{Inotfinal}) becomes
\be
I(x,y,z) =Ê \frac{(\m^2)^{4 - d}}{(4 \pi)^d} \,ÊÊ 
\D^{\frac{d}{2} - \frac32} \,\frac{ 2 \pi \G(5 - d)}{(4-d) (d - 2) (d - 3)} 
+ \G' \, \big( G(\frac{c}{2} - x) +Ê G(\frac{c}{2} - y)
+ G(\frac{c}{2} - z)Ê \big)~,
\label{final}
\ee
which can be caste in the form
\be
I(x,y,z) = - I(\D,0,0) \sin \frac{\pi d}{2} + \G' \, \big( G(\frac{c}{2} - x)
+Ê G(\frac{c}{2} - y) + G(\frac{c}{2} - z) \big)
\ee
using the Gamma function identities $z \G(z) = \G(z+1)$ and
$$ \G(\frac{d}{2} - 1) \G(2 - \frac{d}{2}) = - \, \frac{\pi}{\sin \frac{\pi d}{2}} .$$
This is the result presented in \cite{FJJ}.

For the purposes of examining divergences, 
we also present divergent terms in the $\e$ expansion of $I(x,y,z)$ 
in the limit $d \rightarrow 4.$ 
Two cases are required in this paper: (i)Ê $x, y$ and $z$ all nonzero with $\D>0;$ 
and (ii) $x=0$ and $y$ and $z$ nonzero with $\D<0.$ 
The divergent terms in the $\e$ expansion are not sensitive to the sign of $\D,$ 
and can be read, e.g., from  \cite{FJJ}:
\bea
&& (4 \pi)^4 \, I_{\rm div}(x,y,z) = - \, \frac{1}{2 \e^2} \, (x+y+z) \nonumber \\
&& \qquad + \frac{1}{\e} \, \Big[ x \ln x + y \ln y + z \ln z + 
(\g - \frac32 - \ln 4 \pi \m^2 ) (x + y + z) \Big] 
~,
\label{e-expan1}
\eea
and hence
\bea
&&(4 \pi)^4 \, I_{\rm div}(0,y,z) = - \, \frac{1}{2 \e^2} \, (y+z) \nonumber \\
&&\qquad + \frac{1}{\e} \, \left[Ê y \ln y + z \ln z + (\g - \frac32 - \ln 4 \pi \m^2 ) (y + z) \right] ~.
\label{e-expan2}
\eea
The finite part of $I(x,y,z)$ is sensitive to the sign of $\D$, and it  
is discussed in detail, e.g. in 
\cite{FJJ,DT,DST,CCLR}. For all-order epsilon expansion, see also 
\cite{DK}.

The results for  sunset integrals in  \cite{FJJ}
have been used by many authors for two-loop  calculations of effective potentials in 
various field theories including 
the Standard Model \cite{FJJ}, the Minimal Supersymmetric Standard Model
\cite{EZ,Martin}, and 
also in non-renormalizable supersymmetric theories \cite{GNN}.

\sect{Group-theoretical relations}

In this appendix, we  describe
the $SU(N)$ conventions adopted in this paper.
Lower-case Latin letters from the middle of the alphabet, 
$i,j,\dots$, 
are  used to denote the matrix elements in the fundamental
representation. 
We also set $i=(0,\un{ i})=0,1,\dots, N-1 $.
A generic element of the Lie algebra $su(N)$ is 
\be
u = u^I \, H_I + u^{ij} \, E_{ij} \equiv u^a \,T_a~,  
\qquad i \neq j ~. 
\label{generic}
\ee
We choose a Cartan-Weyl basis 
to consist of the elements:
\be 
H_I = \{ H_0, H_{\un{I}}\}~, \quad  \un{I} = 1,\dots, N-2~, 
\qquad \quad E_{ij}~, \quad i\neq j~. 
\label{C-W}
\ee 
The basis elements 
 defined as matrices in  the fundamental representation are   \cite{KM3,KM4}, 
\bea
(E_{ij})_{kl} &=& \d_{ik}\, \d_{jl}~, \non \\
(H_I)_{kl} &=& \frac{1}{\sqrt{(N-I)(N-I-1)} }
\Big\{ (N-I)\, \d_{kI} \, \d_{lI} - 
\sum\limits_{i=I}^{N-1} \d_{ki} \, \d_{li} \Big\} ~.
\label{C-W-2}
\eea
They satisfy 
\be
{\rm Tr} (H_I\,H_J) = \d_{IJ}\ , \ \ \ \ 
{\rm Tr} (E_{ij}\,E_{kl}) = \d_{il}\,\d_{jk}\ , \ \ \ \   
{\rm Tr} (H_I \,E_{kl}) =0 \ . \ee

\end{appendix}

\small{

}


\begin{thebibliography}{66}

\bi{old}
A.~Parkes and P.~C.~West,
  ``Finiteness in rigid supersymmetric theories,''
  Phys.\ Lett.\ B {\bf 138}, 99 (1984);
  ``Three-loop results in two-loop finite supersymmetric gauge theories,''
  Nucl.\ Phys.\ B {\bf 256}, 340 (1985);
 P.~C.~West,
  ``The Yukawa beta function in N=1 rigid supersymmetric theories,''
  Phys.\ Lett.\  B {\bf 137}, 371 (1984);
D.~R.~T.~Jones and L.~Mezincescu,
  ``The chiral anomaly and a class of two-loop 
  finite supersymmetric gauge theories,''
  Phys.\ Lett.\ B {\bf 138}, 293 (1984);
S.~Hamidi, J.~Patera and J.~H.~Schwarz,
  ``Chiral two-loop finite supersymmetric theories,''
  Phys.\ Lett.\  B {\bf 141}, 349 (1984);
D.~R.~T.~Jones and A.~J.~Parkes,
  ``Search for a three-loop finite chiral theory,''
  Phys.\ Lett.\ B {\bf 160}, 267 (1985);
  A.~V.~Ermushev, D.~I.~Kazakov and O.~V.~Tarasov,
  ``Finite N=1 supersymmetric grand unified theories,''
  Nucl.\ Phys.\  B {\bf 281}, 72 (1987);
 D.~I.~Kazakov,
  ``Finite N=1 SUSY gauge field theories,''
  Mod.\ Phys.\ Lett.\  A {\bf 2}, 663 (1987).

\bibitem{JJN}
  I.~Jack, D.~R.~T.~Jones and C.~G.~North,
  ``$N=1$ supersymmetry and the three-loop anomalous dimension 
  for the chiral superfield,''
  Nucl.\ Phys.\ B {\bf 473}, 308 (1996)
  [hep-ph/9603386].

\bibitem{LS}
  R.~G.~Leigh and M.~J.~Strassler,
  ``Exactly marginal operators and duality in four-dimensional N=1
  supersymmetric gauge theory,''
  Nucl.\ Phys.\ B {\bf 447}, 95 (1995) [hep-th/9503121].

\bibitem{LM}
  O.~Lunin and J.~Maldacena,
  ``Deforming field theories with U(1) x U(1) global symmetry 
and their gravity  duals,''
  JHEP {\bf 0505}, 033 (2005)
  [hep-th/0502086].

\bibitem{FG}
  D.~Z.~Freedman and U.~G\"ursoy,
  ``Comments on the beta-deformed N = 4 SYM theory,''
  JHEP {\bf 0511}, 042 (2005)
  [hep-th/0506128].

\bibitem{PSZ}
  S.~Penati, A.~Santambrogio and D.~Zanon,
   ``Two-point correlators in the beta-deformed N = 4 SYM at the next-to-leading
  order,''
  JHEP {\bf 0510}, 023 (2005)
  [hep-th/0506150].

\bibitem{RSS}
  G.~C.~Rossi, E.~Sokatchev and Y.~S.~Stanev,
  ``New results in the deformed N = 4 SYM theory,''
  Nucl.\ Phys.\ B {\bf 729}, 581 (2005)
  [hep-th/0507113]. 

\bibitem{MPSZ}
  A.~Mauri, S.~Penati, A.~Santambrogio and D.~Zanon,
  ``Exact results in planar N = 1 superconformal Yang-Mills theory,''
  JHEP {\bf 0511}, 024 (2005)
  [hep-th/0507282].
  
\bibitem{KT}
  S.~M.~Kuzenko and A.~A.~Tseytlin,
  ``Effective action of beta-deformed N = 4 SYM theory and AdS/CFT,''
  Phys.\ Rev.\ D {\bf 72}, 075005 (2005)
  [hep-th/0508098].

\bibitem{MPPSZ}
  A.~Mauri, S.~Penati, M.~Pirrone, A.~Santambrogio, D.~Zanon, 
   ``On the perturbative chiral ring for marginally deformed N = 4 SYM
  theories,''
  JHEP {\bf 0608}, 072 (2006)
  [hep-th/0605145].

\bibitem{EMPS}
  F.~Elmetti, A.~Mauri, S.~Penati, A.~Santambrogio and D. Zanon,
   ``Conformal invariance of the planar beta-deformed N = 4 SYM theory 
requires  beta real,''
JHEP {\bf 0701}, 026 (2007) [hep-th/0606125].

\bibitem{RSS2}
  G.~C.~Rossi, E.~Sokatchev and Y.~S.~Stanev,
   ``On the all-order perturbative finiteness of the deformed N = 4 SYM
  theory,'' Nucl.\ Phys.\ B {\bf 754}, 329 (2006) [hep-th/0606284].

\bibitem{AKS}
  S.~Ananth, S.~Kovacs and H.~Shimada,
  ``Proof of all-order finiteness for planar beta-deformed Yang-Mills,''
  JHEP {\bf 0701}, 046 (2007)
  [hep-th/0609149].

\bibitem{DH}
  N.~Dorey and T.~J.~Hollowood,
  ``On the Coulomb branch of a marginal deformation of N = 4 SUSY Yang-Mills,''
  JHEP {\bf 0506}, 036 (2005)
  [hep-th/0411163].

\bibitem{FJ}
Ê C.~Ford and D.~R.~T.~Jones,
Ê ``The effective potential and the differential equations method 
for FeynmanÊ integrals,''
Ê Phys.\ Lett.\ B {\bf 274}, 409 (1992) 409
Ê [Erratum-ibid.\ B {\bf 285}, 399 (1992)].
Ê 

\bibitem{FJJ}
  C.~Ford, I.~Jack and D.~R.~T.~Jones,
  ``The Standard model effective potential at two loops,''
  Nucl.\ Phys.\ B {\bf 387}, 373 (1992)
  [Erratum-ibid.\ B {\bf 504}, 551 (1997)]
  [hep-ph/0111190].

\bibitem{DT}
  A.~I.~Davydychev and J.~B.~Tausk,
   ``Two-loop self-energy diagrams with different masses and the momentum
  expansion,''
  Nucl.\ Phys.\ B {\bf 397}, 123 (1993).

\bibitem{DST}
  A.~I.~Davydychev, V.~A.~Smirnov and J.~B.~Tausk,
  ``Large momentum expansion of two-loop self-energy 
diagrams with arbitrary  masses,''
  Nucl.\ Phys.\ B {\bf 410}, 325 (1993)
  [hep-ph/9307371].

  \bibitem{CCLR}
  M.~Caffo, H.~Czyz, S.~Laporta and E.~Remiddi,
  ``The master differential equations for the 2-loop sunrise selfmass
  amplitudes,''
  Nuovo Cim.\ A {\bf 111}, 365 (1998)
  [hep-th/9805118].

\bibitem{T}
  O.~V.~Tarasov,
  ``Generalized recurrence relations for two-loop propagator integrals with
  arbitrary masses,''
  Nucl.\ Phys.\  B {\bf 502}, 455 (1997)
  [hep-ph/9703319].

\bibitem{NP}
  V.~Niarchos and N.~Prezas,
  ``BMN operators for N = 1 superconformal Yang-Mills theories and associated
  string backgrounds,''
  JHEP {\bf 0306}, 015 (2003)
  [hep-th/0212111].

\bibitem{KM}
  S.~M.~Kuzenko and I.~N.~McArthur,
   ``On the background field method beyond one loop: A manifestly covariant
  derivative expansion in super Yang-Mills theories,''
  JHEP {\bf 0305}, 015 (2003)
  [hep-th/0302205].

\bibitem{KM2}
  S.~M.~Kuzenko and I.~N.~McArthur,
  ``Low-energy dynamics in N = 2 super QED: Two-loop approximation,''
  JHEP {\bf 0310}, 029 (2003)
  [hep-th/0308136].

\bibitem{KM3}
  S.~M.~Kuzenko and I.~N.~McArthur,
   ``On the two-loop four-derivative quantum corrections in 4D N = 2
  superconformal field theories,''
  Nucl.\ Phys.\ B {\bf 683}, 3 (2004)
  [hep-th/0310025].

\bibitem{KM4}
  S.~M.~Kuzenko and I.~N.~McArthur,
  ``Relaxed super self-duality and N = 4 SYM at two loops,''
  Nucl.\ Phys.\ B {\bf 697}, 89 (2004)
  [hep-th/0403240].

\bibitem{GGRS}
S.~J.~Gates, M.~T.~Grisaru, M.~Ro\v{c}ek and W.~Siegel,
{\it Superspace, Or One Thousand 
and One Lessons in Supersymmetry},
Benjamin/Cummings, 1983 [hep-th/0108200].
  
\bibitem{OW}
B.~A.~Ovrut and J.~Wess,
``Supersymmetric $R_\x$ gauge and radiative symmetry breaking,''
Phys.\ Rev.\ D {\bf 25} (1982) 409;
N.~Marcus, A.~Sagnotti and W.~Siegel,
  ``Ten-dimensional supersymmetric Yang-Mills theory in terms of
  four-dimensional superfields,''
  Nucl.\ Phys.\  B {\bf 224}, 159 (1983);
P.~Binetruy, P.~Sorba and R.~Stora,
``Supersymmetric S covariant $R_\x$ gauge,''
Phys.\ Lett.\ B {\bf 129} (1983) 85.


\bibitem{BBP}
A.~T.~Banin, I.~L.~Buchbinder and N.~G.~Pletnev,
``Low-energy effective action in N = 2 super 
Yang-Mills theories on  non-abelian background,''
Phys.\ Rev.\ D {\bf 66}, 045021 (2002) 
[hep-th/0205034];
``One-loop effective action for N = 4 SYM theory 
in the hypermultiplet  sector: Leading low-energy 
approximation and beyond,''
Phys.\ Rev.\ D {\bf 68}, 065024 (2003) 
[hep-th/0304046].

\bibitem{BKT}
  I.~L.~Buchbinder, S.~M.~Kuzenko and A.~A.~Tseytlin,
  ``On low-energy effective actions in N = 2,4 superconformal theories in  four
  dimensions,''
  Phys.\ Rev.\  D {\bf 62}, 045001 (2000)
  [hep-th/9911221].
    
\bibitem{Siegel}
  W.~Siegel,
  ``Supersymmetric dimensional regularization via dimensional reduction,''
  Phys.\ Lett.\  B {\bf 84}, 193 (1979).


\bibitem{Kotikov}
  A.~V.~Kotikov,
  ``Differential equations method: New technique for massive Feynman diagrams
  calculation,''  Phys.\ Lett.\  B {\bf 254}, 158 (1991);
``Differential equations method: The calculation of vertex type Feynman
  diagrams,''
  Phys.\ Lett.\  B {\bf 259}, 314 (1991).

\bi{CH} R. Courant and D. Hilbert, {\it Methods of Mathematical Physics},
Vol. II, Wiley-VCH, 1962.

\bibitem{CT}
 F.~V.~Tkachov,
  ``A theorem on analytical calculability of four loop renormalization group
  functions,''  Phys.\ Lett.\  B {\bf 100}, 65 (1981);
 K.~G.~Chetyrkin and F.~V.~Tkachov,
  ``Integration by parts: The algorithm to calculate beta functions in 4
  loops,''
  Nucl.\ Phys.\  B {\bf 192}, 159 (1981).
 
\bibitem{Broadhurst:1993mw}
  D.~J.~Broadhurst, J.~Fleischer and O.~V.~Tarasov,
  ``Two-loop two-point functions with masses: Asymptotic expansions and Taylor
  series, in any dimension,''
  Z.\ Phys.\ C {\bf 60}, 287 (1993)
  [hep-ph/9304303].

\bi{PBM} A. P. Prudnikov, Yu. A. Brychkov and O. I. Marichev, 
{\it Integrals and Series. Volume 3: More Special Functions},
Gordon and Breach, New York, 1990. 

\bi{Levin} L. Levin, {\it Polylogarithms and Associated Functions},
North Holland, New York, 1981.

\bi{DK} 
A. I. Davydychev, M. Yu. Kalmykov
"New results for the epsilon expansion of certain one, two and three loop Feynman diagrams",
Nucl.Phys.B {\bf 605} (2001) 266  [hep-th/0012189].


\bibitem{EZ}
  J.~R.~Espinosa and R.~J.~Zhang,
  ``Complete two-loop dominant corrections to the mass of the lightest  CP-even
  Higgs boson in the minimal supersymmetric standard model,''
  Nucl.\ Phys.\ B {\bf 586}, 3 (2000)
  [hep-ph/0003246]. 

\bibitem{Martin}
  S.~P.~Martin,
  ``Two-loop effective potential for the minimal supersymmetric standard
  model,'' Phys.\ Rev.\  D {\bf 66}, 096001 (2002)
  [hep-ph/0206136].

\bibitem{GNN}
  S.~Groot~Nibbelink and T.~S.~Nyawelo,
   ``Two loop effective Kaehler potential of (non-) renormalizable 
supersymmetric  models,''
  JHEP {\bf 0601}, 034 (2006)
  [hep-th/0511004].


\end{thebibliography}
\end{document}